\documentclass[10pt]{article}

\usepackage{epsfig}
\usepackage{cite}
\usepackage{natbib}
\usepackage{amsmath}
\usepackage{amsthm}
\usepackage{easybmat}
\usepackage{float}
\usepackage{lmodern}
\usepackage{lscape,epsfig}
\usepackage{amssymb,amsmath}
\usepackage{multirow}
\usepackage{graphicx}
\usepackage{adjustbox}
\usepackage{bm}
\usepackage{ifxetex,ifluatex}
\usepackage[table,xcdraw]{xcolor}
\usepackage{rotating}
\usepackage{booktabs} 

\usepackage{algorithm}
\usepackage[noend]{algpseudocode}
\usepackage[]{algpseudocode}
\usepackage{algorithmicx}
\usepackage{amsfonts}

\algnewcommand{\IfThenElse}[3]{
  \State \algorithmicif\ #1\ \algorithmicthen\ #2\ \algorithmicelse\ #3}

\algloopdefx{NFor}[1]{\textbf{for all} #1 \textbf{do}}
\algnewcommand\And{\textbf{and}}
\algnewcommand\Or{\textbf{or }}

\DeclareMathOperator*{\argmin}{arg\,min}

\IfFileExists{microtype.sty}{\usepackage{microtype}}{} 
\usepackage[margin=1in]{geometry}
\ifxetex
  \usepackage[setpagesize=false, 
              unicode=false, 
              xetex]{hyperref}
\else
  \usepackage[unicode=true]{hyperref}
\fi
\hypersetup{breaklinks=true,
            bookmarks=true,
            pdfauthor={},
            pdftitle={Additive model for interactions},
            colorlinks=true,
            citecolor=blue,
            urlcolor=blue,
            linkcolor=magenta,
            pdfborder={0 0 0}}
\usepackage{color}
\usepackage{soul} 

\usepackage{color}
\definecolor{eGreen}{rgb}{.057, .549,.065}

\usepackage{titling}
  \title{Sufficient Dimension Reduction for Interactions}
  \pretitle{\vspace{\droptitle}\centering\huge}
  \posttitle{\par}
  \author{HYUNG G. PARK$^{a\ast}$, THADDEUS TARPEY$^{a}$, EVA PETKOVA$^{a}$,  
R. TODD OGDEN$^{b}$\\[4pt]
\textit{$^{a}$ Division of Biostatistics, Department of Population Health,
New York University} \\ 
\textit{$^{b}$ Department of Biostatistics, Columbia University}
\\[2pt]
} 
  \preauthor{}\postauthor{}
\date{ \vspace*{-30pt} }

\numberwithin{equation}{section}
\theoremstyle{plain}
\newtheorem{theorem}{Theorem}[section]
\newtheorem{proposition}{Proposition}[section]
\newtheorem{corollary}{Corollary}[section]

\newtheorem{definition}{Definition}
\newtheorem{remark}{Remark}[section]

\begin{document}
\maketitle
\footnotetext{To whom correspondence should be addressed; parkh15@nyu.edu}

\section*{Abstract} 

Dimension reduction lies at the heart of many statistical methods. In regression, dimension reduction has been linked to the notion of sufficiency whereby the relation of the response to a set of predictors is explained by a lower dimensional subspace in the predictor space. In this paper, we consider the notion of a  dimension reduction in regression on subspaces that are sufficient to explain interaction effects between predictors and another variable of interest. The motivation for this work is from precision medicine where the performance of an individualized treatment rule, given a set of pretreatment predictors, is determined by interaction effects.

\noindent
{\it Keywords:}  Precision medicine,  modified covariate method, projection-pursuit regression, single-index models, central mean subspace

\section{INTRODUCTION}\label{sec.intro}

The notion of sufficiency, introduced by Fisher \citep{Fisher1922}, plays a fundamental role in statistics.  A statistic is sufficient if it summarizes all the relevant information in the sample about the parameter of interest.  Sufficiency can be regarded as a form of dimension reduction whereby a sample of size $n$ is reduced to a low-dimension statistic.
Cook \citep[][Section 8.2]{Cook.dimension.reduction} extended the notion of sufficiency to the realm of regression as a dimension reduction concept  \citep[see also,][]{Cook1994, Cook1996, 
Li1991, Li1992, Bura2001, Adragni2009}. Given a set of $p$ covariates $\bm{X} \in \mathbb{R}^p$ and an outcome variable $Y \in \mathbb{R}$, Cook's notion of a sufficient subspace in regression can be summarized as $Y|\bm{X} \sim Y|R(\bm{X})$ where $R: \mathbb{R}^p  \mapsto  \mathbb{R}^q, \; q<p$.

The central subspace, which is denoted by $S_{Y\mid \bm{X}}$, 
is the subspace with the smallest possible dimension $q$ in $\mathbb{R}^p$, such that $Y$ is independent of $\bm{X}$ given $R(\bm{X}) = \bm{B}^\top \bm{X}$ 
for some $p \times q$ matrix $\bm{B}$, $q <p$, 
where the columns of  $\bm{B}$ form a  basis for the subspace \citep{Cook2002}. For comprehensive discussion, see \citep{Cook1998a}. 
Dimension reduction is often aimed at reducing dimensionality for modeling the conditional mean function $\mathbb{E}[Y |\bm{X}]$ alone, while leaving the rest of the distribution $Y | \bm{X}$ as the ``nuisance parameter.'' 
For this case, 
Cook and Li \citep{Cook2002} introduced the central mean subspace, denoted as $S_{\mathbb{E}[Y\mid \bm{X}]}$, 
defined to be the smallest subspace,  
$\mbox{span}(\bm{B})$ for some basis matrix $\bm{B}$, sufficient to model the conditional mean $\mathbb{E}[Y |\bm{X}]$. 

In this paper, we extend the notion of a sufficient subspace in regression with an outcome variable $Y$ when our interest is in the interaction effect between
 the vector of covariates $\bm{X}$ and  
another variable $A \in \mathcal{A}$. 
This paper considers the case when $A$ is a discrete random variable on a space  
$\mathcal{A} = \{1,\ldots, L\}$,   i.e., there are $L$ possible levels for the random variable $A$;  
however, the notion will also be extended  to a continuous compact interval $\mathcal{A} \subset \mathbb{R}$. 
The primary focus is on reducing the dimension of $\bm{X}$ to model the effects of interactions between 
$\bm{X}$ and $A$ on $Y$. 
The motivation for this work is in the context of precision medicine, 
where we seek to optimize an individualized treatment rule that assigns a treatment to
each patient according to  the patient's specific characteristics,  
in the hope of improving efficacy of treatments and lowering medical cost. 
Typically, individual-specific medical/clinical characteristics are represented by a vector of covariates $\bm{X}$ measured before treatment assignment, 
and treatment condition can be represented by the variable $A$. 
An optimal individualized treatment rule is solely determined by 
the $\bm{X}$-by-$A$ interaction effects on $Y$  \citep[e.g., see][]{QianAndMurphy}. 
Therefore, a sufficient reduction subspace for $\bm{X}$ 
in this setting will typically be defined in terms of 
a subspace sufficient to model 
the $\bm{X}$-by-$A$ 
interaction effect,  whereas the pure main effect for $\bm{X}$ on $Y$ 
will be viewed as a ``nuisance'' effect. 

In this paper, 
we define a sufficient dimension reduction subspace 
for $\bm{X}$ in terms of a parsimonious characterization of the $\bm{X}$-by-$A$ interaction effect available from $\mathbb{E}[Y | \bm{X},A]$, and we
introduce a semiparametric framework for producing a basis $\bm{B}$ for
such sufficient subspace. 
The proposed framework of modeling the $\bm{X}$-by-$A$  interactions 
takes the  linear model based approaches  \citep[e.g.,][]{MCA,LU.2011, A.learning.Shi2016, A.learning.Shi2018,A.learning.Jeng2018} 
as its special cases. 
Luo, et al. \citep{Luo2019} considered sufficient dimension reduction to estimate a lower dimensional linear combination of $\bm{X}$ that is sufficient to model the regression causal effect, 
defined as  the mean difference in the potential outcomes 
\citep{Rubin1974} 
conditional on $\bm{X}$ \citep[see also,][]{Luo2017}, when the treatment variable $A$ is binary-valued. 
Our framework, instead, focuses on the interaction between treatment and covariates, allows  $L$ treatment levels, and  it can be easily modified to incorporate the case where $A$ is defined on a continuum.

\section{SUFFICIENT REDUCTION FOR INTERACTIONS}\label{sec.sufficient.reduction}

\subsection{PRELIMINARIES}\label{prelim}

Our approach to the sufficient reduction for interactions is to express the conditional expectation function $\mathbb{E}[Y|\bm{X}, A]$ 
in terms of a  main effect for $\bm{X}$ and a $\bm{X}$-by-$A$ interaction effect.
Consider the following decomposition of the conditional expectation:
\begin{equation} \label{the.decomposition}
\mathbb{E}[Y|\bm{X}, A] \ = \ \mu(\bm{X}) + g(\bm{X}, A),
\end{equation}
where the first  term $\mu(\bm{X})$ does not depend on $A$ 
and 
only the second term $g(\bm{X},A)$ is a function of $A$. 
 Under representation (\ref{the.decomposition}), 
the marginal effect of 
 $\bm{X}$ on $Y$ 
 is expressed as: 
\begin{eqnarray*}
 \mathbb{E}[Y \mid \bm{X}]
&=&
\mathbb{E}[\mathbb{E}[Y \mid \bm{X}, A] \mid \bm{X}]\\
& =& \mathbb{E}[\mu(\bm{X})+ g(\bm{X},A) \mid \bm{X}]\\
&=&\mu(\bm{X}) +  \mathbb{E}[ g(\bm{X}, A) \mid \bm{X} ] 
\end{eqnarray*}
In what follows, 
for the identifiability of decomposition (\ref{the.decomposition}), 
we will set 
\begin{equation} \label{the.identifiability.condition}
\mathbb{E}[ g(\bm{X}, A) | \bm{X} ] = 0.
\end{equation}
The condition (\ref{the.identifiability.condition}) implies that,  in (\ref{the.decomposition}), the first term 
$\mu(\bm{X})  = \mathbb{E}[Y | \bm{X}]$ 
represents the $\bm{X}$ marginal effect,   
and the second term 
$g(\bm{X}, A)$ represents the ``pure'' $\bm{X}$-by-$A$ 
 interaction effect.   
Throughout the paper, we write $\bm{\Sigma} = \mbox{var}(\bm{X})$, and assume an additive mean zero noise with finite variance.

\subsection{CENTRAL MEAN SUBSPACE}

For a discrete treatment space $\mathcal{A}=  \{1, \dots, L\}$ with  $L$ available treatments, 
a treatment decision function, $\mathcal{D}(\bm{X}):  \mathbb{R}^p \mapsto \mathcal{A}$, mapping each individual's pretreatment covariates  $\bm{X} \in \mathbb{R}^p$ to one of the $L$ treatment options,   defines an individualized treatment rule \citep{Murphy, Robins, Zhang.2012, Cai, QianAndMurphy} for s single decision time point. 
The average outcome when all individuals are treated according to such rule is referred to as  the ``value'' $(V)$  of the individualized treatment rule \citep{QianAndMurphy}, which can be expressed as $V(\mathcal{D}) = \mathbb{E}[ \mathbb{E}[Y | \bm{X}, A=\mathcal{D}(\bm{X}) ] ]$. 
Without loss of generality, if we assume a larger value of $Y$ is desirable, then it is straightforward to verify that the optimal individualized treatment rule, $\mathcal{D}^{opt}$, which results in the largest value  $V(\mathcal{D}^{opt})$, is of  the form: 
\begin{equation} \label{dopt}
\mathcal{D}^{opt}(\bm{X}) \ = \  \operatorname*{arg\,max}_{a \in \mathcal{A}} \ \mathbb{E}\big[Y | \bm{X}, A=a \big],
\end{equation}
i.e., the optimal individualized treatment rule assigns a treatment to an individual patient based on the highest expected quality treatment given $\bm{X}$.

We will cast the notion of sufficient reduction for 
 $\bm{X}$-by-$A$ interaction effects 
 under the general representation (\ref{the.decomposition}). 
We define a {\em contrast vector} $\bm{c} =( c_1, \ldots,  c_L )^\top \in \mathbb{R}^L$ as 
a vector such that $\sum_{a=1}^L{c_a}=0$ (zero-sum constraint) 
and  
$ \bm{c} \ne ( 0, \ldots,  0)^\top$, i.e., 
$\bm{c}$ is not a vector of all zeros  (to avoid the trivial case). 

\begin{definition} \label{definition.1}
 For an arbitrary contrast vector $\bm{c} =( c_1, \ldots,  c_L )^\top$,
we define the mean  contrast function  of $\bm{X}$, as the following linear combination: 
\begin{equation} \label{treatment.contrast}
\mathcal{C}(\bm{X};  {\bm c}) 
:= \sum_{a=1}^L c_a  \mathbb{E}\left[ Y |  \bm{X} , A= a \right]. 
\end{equation}  
 \end{definition}

The mean contrast function (\ref{treatment.contrast}) 
is a transformation of the function 
$\mathbb{E}\left[ Y | \bm{X}, A \right]$ in  (\ref{the.decomposition}) 
from its original domain $(\bm{X},A)$ 
to the new domain $(\bm{X}, \bm{c})$. 
The condition $\sum_{a=1}^L c_a  =0$ imposed on $\bm{c}$ makes the $\bm{X}$ marginal effect $\mu(\bm{X})$  in the general model (\ref{the.decomposition}) drop out from 
$\mathcal{C}(\bm{X}; {\bm c})$ in (\ref{treatment.contrast}). 
As a result, 
the mean contrast function  (\ref{treatment.contrast}), for any $\bm{c}$, is independent of the marginal effect $\mu(\bm{X})$ in (\ref{the.decomposition}).

In a treatment context, $\mathcal{C}(\bm{X};  \bm{c})$  is a measure comparing individualized efficacies of treatments for  a given contrast $\bm{c}$, 
as a function of the pretreatment covariates $\bm{X}$. 
For example, if  $L=2$, 
the optimal individualized treatment rule defined in (\ref{dopt}) 
 is determined by the sign of $\mathcal{C}(\bm{X};  \bm{c})$ 
 when $c_1 = 1$ and $c_2= -1$, 
 and  (\ref{treatment.contrast})  is reduced to the case studied by Luo, et al. \citep{Luo2019}.

In this paper, we consider a lower dimensional transformation of 
$\bm{X}$ that is sufficient to recover the mean contrast functions $\mathcal{C}(\bm{X};  \bm{c})$ in 
 (\ref{treatment.contrast}), for any contrast vector $\bm{c}$. 
 
  \begin{definition} \label{definition.2}
Let $\bm{B}$ denote a $p\times q$ matrix with full column rank. The transformation
$R(\bm{X}) = \bm{B}^\top \bm{X}$ is said to be a sufficient dimension reduction for  
$\bm{X}$-by-$A$  
  interactions,  if 
\begin{equation} \label{treatment.contrast.reduction}
\mathcal{C}(\bm{X};  \bm{c} ) 
=\mathcal{C}(\bm{B}^\top \bm{X};  \bm{c} ) = \sum_{a=1}^L c_a   g_a(\bm{B}^\top \bm{X}) 
 \end{equation}
 for any contrast vector $\bm{c}$, 
where  the functions 
 $\{ g_a \}_{a \in \mathcal{A}}$
are unspecified functions 
associated with each level of $A \in \mathcal{A}$ defined on $\bm{B}^\top \bm{X} \in \mathbb{R}^q$. 
The column space of $\bm{B}$ will be called a sufficient reduction subspace for $\bm{X}$-by-$A$ interactions.
 \end{definition}

 For any $p\times q$  matrix $\bm{B} = (\bm{\beta}; \ldots; {\bm{\beta}}_q )$ 
 satisfying (\ref{treatment.contrast.reduction}) 
 and  any $p \times p$ nonsingular matrix $\bm{\eta}$,  $\bm{\eta} \bm{B}$ still satisfies  (\ref{treatment.contrast.reduction})  when the 
 $\{ g_a \}_{a \in \mathcal{A}}$
 are adjusted accordingly. 
A further constraint on $\bm{B}$ is needed for an identifiable parametrization. 
To remove trivial ambiguity, let us define a set of $p \times q $ matrices, denoted as $\Theta_q$, that have a positive first nonzero entry and
consists of $q$ distinct orthonormal vectors;
%
in (\ref{treatment.contrast.reduction}), without loss of generality, we assume $\bm{B} \in \Theta_q$.

 As the notion of sufficiency (\ref{treatment.contrast.reduction}) is based on the contrast function $\mathcal{C}(\bm{X};  \bm{c} )$ 
 that is independent of $\mu(\bm{X})$ in   (\ref{the.decomposition}), 
 we can formalize a minimally sufficient dimension reduction in $\bm{X}$ specifically for the term $g(\bm{X},A)$ in   (\ref{the.decomposition}).

  \begin{definition} \label{definition.3}
A sufficient reduction subspace for interactions is said to be minimal, if the dimension of its span is less than or equal to the dimension of the span of any other
  sufficient reduction subspace for interactions. We denote the minimally sufficient reduction subspace (also called the central mean subspace) for $\bm{X}$-by-$A$ interactions as
 $S_{\mathcal{C}| \bm{X}}$, and  $\mbox{dim}(S_{\mathcal{C}| \bm{X}})$ will denote its dimension. 
 \end{definition}

The central mean subspace of \cite{Cook2002} refers to the minimally sufficient subspace in $\mathbb{R}^p$  associated with the mean response function.  
The subspace $S_{\mathcal{C}| \bm{X}}$ is a special case of  the central mean subspace for the mean function (\ref{the.decomposition})  
in which only the interaction term  
$g(\bm{X},A)$ is considered for dimension reduction. 
 We assume that the central mean subspace for interactions, 
$S_{\mathcal{C}| \bm{X}}$,  uniquely exists throughout this article.  
 The uniqueness of the central mean subspace is guaranteed under fairly general conditions  
 \citep{Cook2002, Luo2019, Yin2008}; for example, it is guaranteed when the domain of $\bm{X}$ is open and convex.

\begin{remark}
If there exists a $p \times q$ dimension reduction matrix $\bm{B}$ with $q < p$  
such that $\mathcal{C}(\bm{X};  \bm{c}) = \mathcal{C}(\bm{B}^\top\bm{X};  \bm{c})$, the 
corresponding transformation is sufficient (for interactions) based on Definition \ref{definition.2}, but this need not be a minimal sufficient reduction. For example,
if $\bm{X} = (X_1, X_2)^\top$ 
and $\mathbb{E}[Y | \bm{X}, A=a] = X_1+\gamma_a X_2$ $(a \in \mathcal{A})$ 
 (for $\gamma_a \in \mathbb{R}$ and $\mbox{var}(\gamma_A)>0$), then the $2 \times 2$ identity matrix 
 $\bm{B} = 
\begin{pmatrix}
1 & 0  \\
0 & 1 \\
\end{pmatrix}
$  
  corresponds to a sufficient reduction, but the minimal sufficient reduction is determined by the vector $\bm{B} = (0,1)^\top \in \Theta_1$, since 
  the effect of $A$ is a function only of $X_2$ and  does not depend on $X_1$. 
\end{remark}

Dimension reduction using a minimal number, $\mbox{dim}(S_{\mathcal{C}| \bm{X}})$, of directions  is important for interpretability and  
parsimonious parametrization,  
and allows a more accurate estimation. 
In practice, 1-dimensional reductions often suffice in capturing pertinent interaction effects and are typically of primary interest.
Examples of 1-dimensional reductions include performing a regression with 
a linear model that focuses on a single 
vector of coefficients \citep[e.g.,][]{MCA,GEM.Petkova} 
  and its 
 semiparametric generalization with a set of flexible link functions, 
 a single-index model  with treatment level-specific link functions \citep{SIMML}. 
%
In the remainder of the paper, 
we introduce a semiparametric regression framework for approximating the minimally sufficient subspace $S_{\mathcal{C}| \bm{X}}$ for $\bm{X}$-by-$A$ interactions and build connections to other linear model-based approaches as its special cases.

\subsection{THE MODEL}

Motivated by the notion of sufficiency (\ref{treatment.contrast.reduction}), 
we posit that the $\bm{X}$-by-$A$  interaction effect  from 
 $\mathbb{E}\left[Y | \bm{X}, A\right]$  in (\ref{the.decomposition})
has an intrinsic $q$-dimensional structure 
with some dimension reduction matrix $\bm{B}_0 \in \Theta_q$ of rank-$q$: 
\begin{equation} \label{CSIM.model}
\mathbb{E}\left[Y \mid \bm{X}, A=a\right] 
 =  \mu(\bm{X})  
 +  
  g_{0a}\big( \bm{B}_0^\top \bm{X} \big)  \quad (a \in \mathcal{A}). 
\end{equation}
Here $\mu(\bm{X})$ is an unspecified square integrable  function of $\bm{X}$ only, 
and   as in (\ref{the.identifiability.condition}), 
the expected value of the second 
term $ g_{0A}\big( \bm{B}_0^\top \bm{X} \big)$ 
given $\bm{X}$ is zero,  i.e., 
\begin{equation} \label{the.constraint}
\mathbb{E}[ g_{0A}(\bm{B}_0^\top \bm{X}) | \bm{X}] =  0, 
  \end{equation}
for model identifiability. 
Let $\mathcal{H}^{(\bm{B})}$ denote the Hilbert space of measurable functions 
of $\bm{B}^\top \bm{X}$ for each fixed $\bm{B} \in \Theta_q$, 
and, in (\ref{CSIM.model}), we assume $g_{0a}(\bm{B}_0^\top \bm{X}) \in \mathcal{H}^{(\bm{B}_0)}$ $(a \in \mathcal{A})$.  
Only to  simplify the illustration and to suppress the treatment level $a$-specific intercepts,  
 we assume, without loss of generality, $\mathbb{E}[Y | A=a] =0$, 
  i.e., 
the outcome 
$Y$ is centered within each treatment level $a$ $(a \in \mathcal{A})$, and 
 this can be satisfied by  removing the treatment level $a$-specific means  from $Y$.

 \begin{theorem} \label{theorem.1}
 For the mean model of form (\ref{the.decomposition}),  the lower dimensional representation (\ref{CSIM.model}) of the $\bm{X}$-by-$A$ interaction effect term 
 is equivalent to 
the sufficiency of the reduction $R(\bm{X}) = \bm{B}_0^\top \bm{X}$ for the central mean subspace $\mathcal{S}_{C|\bm{X}}$. 
  \end{theorem}

\begin{corollary} \label{corollary.1}
The set of columns of $\bm{B}_0$ in model (\ref{CSIM.model}) is a basis of the central mean subspace $S_{\mathcal{C}|\bm{X}}$. 
\end{corollary}
  
 Theorem~\ref{theorem.1} and 
Corollary~\ref{corollary.1} indicate that if our interest is in the estimation of $S_{\mathcal{C}|\bm{X}}$, 
we  can focus on estimating $\bm{B}_0$ of model (\ref{CSIM.model}).

\subsection{CRITERION} 

Under model  (\ref{CSIM.model}), 
we can view the treatment $a$-specific functions $\{ g_{0a} \}_{a \in \mathcal{A}}$ 
   and  the dimension reduction matrix $\bm{B}_0$  as 
the solution to the following optimization:  
 \begin{equation} \label{the.criterion}
\begin{aligned}
(g_{01},\ldots, g_{0L}, \bm{B}_0)
 \quad = \quad & \underset{ g_a \in \mathcal{H}^{(\bm{B})}, \bm{B} \in \Theta_q }{\text{argmin}}
& &\mathbb{E} \big[ \big(Y -  \mu(\bm{X})  -  g_{A}( \bm{B}^\top \bm{X}) \big)^2 \big]  \\
& \text{subject to} & &\mathbb{E}\left[ g_{A}( \bm{B}^\top \bm{X}) | \bm{X}  \right]  = 0, 
\end{aligned}
\end{equation}
where 
$\mu(\bm{X})$ is the fixed term given from the assumed model (\ref{CSIM.model}).

The first line of the right-hand side  of (\ref{the.criterion}) is 
 \begin{equation*}\label{the.criterion2}
\begin{aligned}
& \argmin_{g_a \in \mathcal{H}^{(\bm{B})}, \bm{B} \in \Theta_q}   
 \mathbb{E}\left[   Y^2  +  \big(g_A(\bm{B}^\top \bm{X})\big)^2   - 2 g_A(\bm{B}^\top \bm{X})Y + 2  g_A(\bm{B}^\top \bm{X}) \mu(\bm{X}) \right]  \\
  =&  \argmin_{g_a \in \mathcal{H}^{(\bm{B})}, \bm{B} \in \Theta_q}  \mathbb{E}\left[    Y^2  +  \big(g_A(\bm{B}^\top \bm{X})\big)^2   - 2 g_A(\bm{B}^\top \bm{X})Y +  2 \mathbb{E}\big[  g_A(\bm{B}^\top \bm{X}) \mu(\bm{X}) | \bm{X} \big] \right] \\ 
   =&  \argmin_{g_a \in \mathcal{H}^{(\bm{B})}, \bm{B} \in \Theta_q}  \mathbb{E}\left[   Y^2  +  \big(g_A(\bm{B}^\top \bm{X})\big)^2   - 2 g_A(\bm{B}^\top \bm{X})Y   \right],  
\end{aligned}
\end{equation*}
where the first equality follows from an application of the iterated expectation rule to condition on $\bm{X}$
 and the second equality follows from the constraint $\mathbb{E}\left[ g_{A}( \bm{B}^{\top}\bm{X}) | \bm{X}  \right]  = 0$ in (\ref{the.criterion}). 
Therefore, representation (\ref{the.criterion}) can be simplified to: 
 \begin{equation} \label{LS4}
\begin{aligned}
(g_{01},\ldots, g_{0L}, \bm{B}_0) \quad = \quad & \underset{g_a \in \mathcal{H}^{(\bm{B})}, \bm{B} \in \Theta_q }{\text{argmin}}
& &\mathbb{E} \big[ \big(Y -  g_{A}( \bm{B}^\top \bm{X}) \big)^2 \big]  \\
& \text{subject to} & &\mathbb{E}\left[ g_{A}( \bm{B}^\top\bm{X}) | \bm{X}  \right]  = 0, 
\end{aligned}
\end{equation}
Representation (\ref{LS4}) of the parameters of interest $(g_{01},\ldots, g_{0L}, \bm{B}_0)$ of the dimension reduction model (\ref{CSIM.model}) is particularly useful 
when the ``nuisance parameter'' $\mu$ is a complicated function, difficult  to specify correctly.

The constrained least squares framework (\ref{LS4}) provides a class of regression approaches to estimating the subspace $S_{\mathcal{C}|\bm{X}} = \mbox{span}(\bm{B}_0)$. 
Specifically, the objective function of the right-hand side of (\ref{LS4}) can be approximated based on a sample $(y_i, a_i, \bm{x}_i)$ $(i=1,\ldots,n$), 
where 
the $a$-specific functions 
$\{ g_{a} \}_{a \in \mathcal{A}}$  are appropriately estimated 
subject to the constraint in  (\ref{LS4}). 
Representation (\ref{LS4}) extends the existing linear approaches to modeling interactions  into a semiparametric framework, as will be illustrated in Sections~\ref{sec.linear.model} and \ref{sec.semipar.model}, 
where we focus on the case of a randomized clinical trial, in which the treatment 
$A \in \{1,\ldots,L\}$ is assigned independently of $\bm{X}$ with some probabilities $(\pi_1,\ldots, \pi_L)$, $\sum_{a=1}^L \pi_a = 1$ and $\pi_a > 0$. 

\begin{remark}
For (\ref{LS4}), 
the constraint $\mathbb{E}\left[ g_{A}( \bm{B}^{\top}\bm{X}) | \bm{X}  \right]  = 0$  imposed on the link-functions $\{ g_{a} \}_{a \in \mathcal{A}}$ 
parallels the constraint $\sum_{a=1}^L c_a = 0$ imposed on $\{ c_a g_{a} \}_{a \in \mathcal{A}}$  
of  (\ref{treatment.contrast.reduction}).
 \end{remark}

\section{THE LINEAR MODEL} \label{sec.linear.model}

Let us first consider 
a classical linear model for the $\bm{X}$-by-$A$ interaction effect 
defined based on a set of  the treatment $a$-specific (length-$p$) coefficient vectors 
$\bm{\eta}_a :=  \bm{\Sigma}^{-1} \mbox{cov}[ Y, \bm{X} | A=a ]$ $(a \in \mathcal{A})$.  
The model is written as:
\begin{equation} \label{linear.model}
\mathbb{E}[Y | \bm{X}, A=a] \  =  \ \tilde{\mu}(\bm{X})  \ + \  {\bm{\eta}}_a^\top  \bm{X}  \quad (a \in \mathcal{A}), 
\end{equation} 
where 
the first term $\tilde{\mu}(\bm{X})$ 
represents an unspecified  main effect of $\bm{X}$ that does not depend on $A$. 
To study the $\bm{X}$-by-$A$ interaction effect in the framework of the dimension reduction model  (\ref{CSIM.model}), 
let us introduce the  $p \times p$  
``dispersion'' matrix of the treatment $a$-specific coefficients $\{ \bm{\eta}_a  \in \mathbb{R}^p\}_{a \in \mathcal{A}}$ 
of model (\ref{linear.model}), 
\begin{equation} \label{between.matrix} 
\bm{H} = \sum_{a=1}^L  \pi_a (\bm{\eta}_a - \bar{\bm{\eta}} )(\bm{\eta}_a - \bar{\bm{\eta}} )^\top, 
\end{equation}
where $\bar{\bm{\eta}} = \mathbb{E}[\bm{\eta}_A] =   \sum_{a=1}^L  \pi_a \bm{\eta}_a \in \mathbb{R}^p$. 
Define 
$\bm{\Xi}:= \left( \bm{\xi}_1;  \ldots; \bm{\xi}_{L-1} \right) \sim
p \times (L-1)$,  
as the matrix consisting of the eigenvectors $(\bm{\xi}_1, \ldots, \bm{\xi}_{L-1})$ of 
the matrix $\bm{H}$  (\ref{between.matrix}) 
associated with the $L-1$ 
leading eigenvalues (there are only $L-1$ nonzero eigenvalues; we assume $p > L-1$). 
The following proposition states that, 
when $\{ \bm{\eta}_a \}_{a \in \mathcal{A}}$  
are distinct, 
$\mbox{span}( \bm{\Xi})$ corresponds to 
the central mean subspace $S_{\mathcal{C}|\bm{X}}$.

\begin{proposition} \label{proposition.1}
Under the linear interaction model  (\ref{linear.model}), 
$\mathcal{C}\left( \bm{X}; \bm{c} \right)= \mathcal{C}\left( \bm{\Xi}^\top \bm{X}; \bm{c}  \right)$ 
for any contrast vector $\bm{c}$, 
and thus 
 $\mbox{span}(\bm{\Xi})$ provides a sufficient reduction for 
  (\ref{treatment.contrast}). 
  Furthermore,  
  $S_{\mathcal{C}|\bm{X}}=  \mbox{span}( \bm{\Xi} )$,  if 
  $\{ \bm{\eta}_a \}_{a \in \mathcal{A}}$   are distinct.  
\end{proposition}
The proof of Proposition~\ref{proposition.1} is in the Appendix.   
If we cast model (\ref{linear.model}) 
under the dimension reduction model 
(\ref{CSIM.model}),  
Proposition~\ref{proposition.1} implies  that  
$\mbox{span}(\bm{\Xi}) = \mbox{span}(\bm{B}_0)$ and $\mbox{dim}(S_{\mathcal{C}|\bm{X}}) = L-1$. 
In the context of optimizing an individualized treatment rule, Proposition~\ref{proposition.1} indicates that one can focus on estimating the eigenvectors $(\bm{\xi}_1, \ldots, \bm{\xi}_{L-1})$ of $\bm{H}$, if the $\bm{X}$-by-$A$ interaction effects are linear (\ref{linear.model}). 
Next, we will describe the estimation of the leading eigenvector $\bm{\xi}_1,$ 
in the optimization framework of 
 (\ref{LS4}).

\subsection{A LINEAR GENERATED EFFECT-MODIFIER (GEM) MODEL } \label{sec.linear.link}  

A useful 1-dimensional approximation to the 
 linear $\bm{X}$-by-$A$ interaction model (\ref{linear.model}) is: 
\begin{equation}  \label{eq1.constrained.simml}
\mathbb{E}[Y \mid \bm{X}, A=a] \ \approx \  \tilde{\mu}(\bm{X}) + \tilde{\gamma}_a \bm{\beta}^\top \bm{X} \quad (a \in \mathcal{A}), 
\end{equation}
for a 1-dimensional (1-D) projection vector $\bm{\beta} \in \Theta_1$  (for identifiability).   
 Model (\ref{eq1.constrained.simml}) can be used to approximate  the basis of the subspace 
$S_{\mathcal{C}| \bm{X}}$, i.e., $\bm{\Xi}$,  based on a  rank-1  
projection determined by $\bm{\beta}$. 
In (\ref{eq1.constrained.simml}), 
the $\bm{X}$-by-$A$ interaction effect term 
$\tilde{\gamma}_a {\bm{\beta}}^\top \bm{X}$ $(a \in \mathcal{A})$ 
captures the variability in $\bm{X}$ related to $A$ 
via a 1-dimensional projection $\bm{\beta}^\top \bm{X}$,   
and its interaction with $A  \in \mathcal{A}$ via the $a$-specific slopes $\tilde{\gamma}_a \in \mathbb{R}$ $(a \in \mathcal{A})$.  
\cite{GEM.Petkova} called the projection $\bm{\beta}^\top \bm{X}$ a {\em generated effect-modifier}, 
as it combines $p$ pretreatment covariates $\bm{X}$ into a single (composite) treatment effect-modifier. 
Model (\ref{eq1.constrained.simml}) 
is useful for visualizing the heterogenous treatment (i.e., variable $A$) effects along the particular ``biosignature'' 
axis  $\bm{\beta}^\top \bm{X}$. 
As in (\ref{linear.model}), the term 
$\tilde{\mu}(\bm{X})$ in (\ref{eq1.constrained.simml}) represents an unspecified main effect of $\bm{X}$.  

Let us cast model (\ref{eq1.constrained.simml}) under (\ref{CSIM.model}) by centering (shifting) 
$\gamma_{a} := \tilde{\gamma}_a  - \bar{\gamma}$ $(a \in \mathcal{A})$, 
where $\bar{\gamma}  := 
 \sum_{a=1}^L \pi_{a} \tilde{\gamma}_a$. 
 The resulting reparametrized (i.e., shifted) model (\ref{eq1.constrained.simml}) is 
\begin{equation} \label{eq2.constrained.simml}
\mathbb{E}[Y \mid  \bm{X}, A=a] \ \approx  \  \mu(\bm{X}) +  \gamma_{a} {\bm{\beta}}^\top  \bm{X} \quad (a \in \mathcal{A}), 
\end{equation} 
subject to the identifiability condition of this particular reparametrization: 
 \begin{equation} \label{linGEM.constraint}
\sum_{a=1}^L \pi_a \gamma_{a}  = 0.  
\end{equation}   
In (\ref{eq2.constrained.simml}), the first term 
$\mu(\bm{X})= \tilde{\mu}(\bm{X}) +   \bar{\gamma}  {\bm{\beta}}^\top \bm{X}$ 
corresponds to the reparametrized  (i.e., shifted) main effect term associated with $\bm{X}$.   
The constraint (\ref{linGEM.constraint}) implies $\mathbb{E}[ \gamma_A \bm{\beta}^\top \bm{X} | \bm{X}]  = 0$ for any arbitrary $\bm{\beta} \in \Theta_1$, 
a special case of the constraint in (\ref{the.criterion}), 
where  the functions $\{ g_a  \in \mathcal{H}^{(\bm{B})} \}_{a \in \mathcal{A}}$
are replaced by 
the 
slopes  $\{ \gamma_a  \in \mathbb{R} \}_{a \in \mathcal{A}}$
and the matrix $\bm{B}$  is replaced by the vector $\bm{\beta}$.

To optimize the interaction effect parameters $\{ \gamma_a \bm{\beta} \}_{a \in \mathcal{A}}$
of the rank-1 approximation model (\ref{eq2.constrained.simml}), 
we employ criterion (\ref{LS4}) and this corresponds to solving: 
 \begin{equation}\label{least squares2}
  \underset{ \gamma_a \in \mathbb{R},\  \bm{\beta} \in \Theta_1}{\text{argmin}}
  \ \mathbb{E} \big[ \big(Y  -  \gamma_{A}{\bm{\beta}}^\top  \bm{X} \big)^2\big],
\end{equation}
 subject to the constraint  (\ref{linGEM.constraint}), 
 where the minimization is over both the slopes $\{ \gamma_a\}_{a \in \mathcal{A}}$ and 
 the vector $\bm{\beta}$. 
The following proposition provides an  explicit  expression of solutions 
$\{ \gamma_a\}_{a \in \mathcal{A}}$ of (\ref{least squares2}) for a fixed $\bm{\beta} \in \Theta_1$. 
\begin{proposition} \label{proposition.2}
For the linear $\bm{X}$-by-$A$ interaction model (\ref{linear.model}), 
the solutions 
$\{ \gamma_a\}_{a \in \mathcal{A}}$ 
of (\ref{least squares2}) 
 for a fixed vector $\bm{\beta}$ 
is given by 
$\gamma_{a} = ( {\bm{\beta}}^\top \bm{\Sigma} {\bm{\beta}})^{-1}  {\bm{\beta}}^\top \bm{\Sigma}  (\bm{\eta}_a - \bar{\bm{\eta}})$ 
 $(a \in \mathcal{A})$, where $\bar{\bm{\eta}} = \sum_{a=1}^L \pi_a \bm{\eta}_a$.  
\end{proposition}
\noindent
The proof of Proposition~\ref{proposition.2} is in the Appendix. 
By  Proposition~\ref{proposition.2}, 
$\mbox{var}(\gamma_A \bm{\beta}^\top \bm{X}) = \mathbb{E}({\gamma_A}^2) \mbox{var}({\bm{\beta}}^\top \bm{X})$ (note, $\mathbb{E}({\gamma_A}) =0$ by (\ref{linGEM.constraint})) 
can be explicitly written as:  
 \begin{equation} \label{the.numerator.method}
 \begin{aligned}
\mbox{var}(\gamma_A \bm{\beta}^\top \bm{X})
&= \sum_{a=1}^L \pi_a \frac{({\bm{\beta}}^\top \bm{\Sigma}  (\bm{\eta}_a - \bar{\bm{\eta}}))^2}{ {\bm{\beta}}^\top \bm{\Sigma} \bm{\beta}}  \\
&= \frac{ {\bm{\beta}}^\top \bm{\Sigma} \left[  \sum_{a=1}^L \pi_a  (\bm{\eta}_a - \bar{\bm{\eta}})  (\bm{\eta}_a - \bar{\bm{\eta}})^\top   \right] \bm{\Sigma}  {\bm{\beta}}  }
{{\bm{\beta}}^\top  \bm{\Sigma} {\bm{\beta}} }   \\ 
&= 
\frac{ {\bm{\beta}}^\top \bm{\Sigma} \bm{H} \bm{\Sigma}  {\bm{\beta}}  }
{{\bm{\beta}}^\top \bm{\Sigma}  {\bm{\beta}} } 
= \frac{ \tilde{\bm{\beta}}^{\top} \bm{\Sigma}^{1/2} \bm{H} \bm{\Sigma}^{1/2}{\tilde{\bm{\beta}}}  }
{ \tilde{\bm{\beta}}^{\top} \tilde{\bm{\beta}} },  
\end{aligned}
\end{equation} 
where $\bm{H}$ is defined in (\ref{between.matrix}). 
In the last equality of (\ref{the.numerator.method}), 
 $\tilde{\bm{\beta}} = \bm{\Sigma}^{1/2} \bm{\beta}$, where 
 $\bm{\Sigma}^{1/2} $ is the symmetric ``square root'' of $\bm{\Sigma}$.  
Minimizing criterion (\ref{least squares2})  over $\bm{\beta} \in \Theta_1$  is equivalent to maximizing  (\ref{the.numerator.method}) over $\bm{\beta} \in \Theta_1$; 
it is clear that (\ref{the.numerator.method})
is maximized 
if 
$\tilde{\bm{\beta}}$  is 
the leading eigenvector of $ \bm{\Sigma}^{1/2} \bm{H} \bm{\Sigma}^{1/2}  =  \bm{\Sigma}^{1/2} \bm{\Xi} \bm{\Lambda} \bm{\Xi}^\top \bm{\Sigma}^{1/2}$, 
in which $\bm{\Lambda}$ is the diagonal matrix  consist of the leading  eigenvalues of $\bm{H}$.  
Thus, the maximizer $\tilde{\bm{\beta}}$ of (\ref{the.numerator.method}) 
is the 
 leading column vector 
 of $\bm{\Sigma}^{1/2} \bm{\Xi}$. 
Since $\bm{\beta} = \bm{\Sigma}^{-1/2} \tilde{\bm{\beta}}$, 
the maximizer $\bm{\beta}$  of  (\ref{the.numerator.method})  is the leading column vector of   $\bm{\Xi}$, i.e., $\bm{\xi}_1$.  
Together with Proposition~\ref{proposition.2}, we have the following proposition for 
model (\ref{eq2.constrained.simml}). 

\begin{proposition} \label{theorem.2}
Under model (\ref{linear.model}), 
the solution $\bm{\beta} \in \Theta_1$  of (\ref{least squares2})   for the approximation model  (\ref{eq2.constrained.simml}) 
is   $\bm{\beta}= \bm{\xi}_1$, the leading eigenvector associated with $\bm{H}$.   
The  corresponding treatment 
 $a$-specific slope is  $\gamma_{a} = ( {\bm{\xi}_1}^\top  \bm{\Sigma}  {\bm{\xi}_1})^{-1}  {\bm{\xi}_1}^\top \bm{\Sigma}  (\bm{\eta}_a - \bar{\bm{\eta}})$ $(a \in \mathcal{A})$.
\end{proposition}

Thus, the criterion 
(\ref{least squares2}) 
produces a vector $(\bm{\xi}_1)$ in the central mean subspace $S_{\mathcal{C}| \bm{X}}$ for interactions.

\subsection{EQUIVALENCE TO THE MODIFIED COVARIATE MODEL} \label{sec.connection.MC}

In the special case of $L=2$ levels (i.e., when $A$ is binary-valued), 
the ``modified covariate'' \citep[][]{MCA} method of modeling the $\bm{X}$-by-$A$ interaction effect 
posits the  model  \citep[see also][]{LU.2011, A.learning.Shi2016, A.learning.Shi2018,A.learning.Jeng2018}: 
 \begin{equation} \label{the.MC}
\mathbb{E}[Y \mid  \bm{X}, A=a ] \ = \mu(\bm{X}) \ +   \bm{\beta}^{\ast \top}  \bm{X} (a +\pi_1 -2)
\quad \quad  
 (a \in \{1,2\}), 
\end{equation}
for some coefficient vector $\bm{\beta}^\ast \in \mathbb{R}^p$, 
where the first term $\mu(\bm{X})$ represents an unspecified main effect of $\bm{X}$ 
(and  $\pi_1=\mbox{pr}(A=1)$).

Taking the unspecified  functions 
$\{ g_{0a} \}_{a \in \mathcal{A}}$  
in (\ref{CSIM.model})  
 to a pre-specified form: 
 \begin{equation} \label{the.MC2}
 g_{0a} (u) = (a +\pi_1 -2) u \quad (a \in \{1,2\})
\end{equation}
with $u =  \bm{\beta}^{\ast \top}  \bm{X}$ 
and taking 
 $\bm{B}_0 =  \bm{\beta}^\ast$,  
 reduces the dimension reduction model (\ref{CSIM.model})   to the modified covariate model (\ref{the.MC}). 
The set of $a$-specific functions $\{ g_{01},g_{02}\}$  in  (\ref{the.MC2})  satisfies the identifiability condition (\ref{the.constraint}) 
of model (\ref{CSIM.model}), i.e., 
$\mathbb{E}[ \bm{\beta}^{\ast \top}  \bm{X} (A+ \pi_1 -2) | \bm{X}] = \bm{\beta}^{\ast \top}  \bm{X} \mathbb{E}[ A+ \pi_1 -2 ] 
 = 0$ (almost surely). This allows us to    
represent the coefficient  $\bm{\beta}^\ast$ of (\ref{the.MC})  based on the optimization framework  (\ref{LS4}): 
\begin{equation} \label{the.MC.criterion}
\bm{\beta}^\ast \ = 
\   \underset{ \bm{\beta} \in \mathbb{R}^p}{\text{argmin}} \
\mathbb{E}\big[ \big(Y - {\bm{\beta}}^\top  \bm{X} (A+ \pi_1 -2)  \big)^2 \big], 
\end{equation} 
without involving the term  
$\mu(\bm{X})$ in model (\ref{the.MC}). 
Based on a sample $(y_i, a_i, \bm{x}_i)$ $(i=1,\ldots,n)$, solving 
an empirical version of 
(\ref{the.MC.criterion}) produces a 
consistent estimator of  ${\bm{\beta}}^{\ast}$, with $\mu(\bm{X})$ in model (\ref{the.MC}) unspecified.

When $L=2$, 
under the assumption of 
the linear $\bm{X}$-by-$A$ interactions (\ref{linear.model}), 
there is an equivalence between optimization (\ref{least squares2}) and the right-hand side of 
(\ref{the.MC.criterion}),  
in terms of the vectors derived from the two optimizations. 
If $L=2$, the subspace $S_{\mathcal{C}|\bm{X}}$ given from  Proposition~\ref{proposition.1} is rank-1, 
 and it is spanned by the eigenvector $\bm{\xi}_1$
associated with the only one non-zero eigenvalue of $\bm{H}$ in (\ref{between.matrix}).  In particular, $\bm{\xi}_1= (\bm{\eta}_2 - \bm{\eta}_1)/ \lVert \bm{\eta}_2 - \bm{\eta}_1 \rVert$  \citep{GEM.Petkova}, 
up to a sign.  
The equivalency follows from Proposition~\ref{theorem.2}  that gives an explicit expression of the minimizer 
 $(\gamma_1, \gamma_2, \bm{\beta})$ of (\ref{least squares2})  in terms of the population parameters in (\ref{linear.model}), and the expression for $\bm{\xi}_1$ available in a closed form.

\begin{proposition} \label{proposition.3}
For the linear $\bm{X}$-by-$A$ interaction model (\ref{linear.model}) with $L=2$, 
the solution
$\bm{\beta}^{\ast} $ of (\ref{the.MC.criterion}) 
satisfies:  
$\bm{\beta}^{\ast}  =  \bm{\xi}_1$, up to a scale constant. That is, 
under  (\ref{linear.model}) with $L=2$, there is an equivalence between  (\ref{least squares2})  and  (\ref{the.MC.criterion}) 
in terms of producing vectors in $S_{\mathcal{C}|\bm{X}} (= \mbox{span}(\bm{\xi}_1))$.  
\end{proposition}

Proposition~\ref{proposition.3} indicates that the modified covariate method
(i.e., the right-hand side of (\ref{the.MC.criterion}))
produces a vector in the subspace $S_{\mathcal{C}|\bm{X}}$  when $L=2$.  
It follows that, 
 in the special case of $L=2$, 
the rank-1 approximation model (\ref{eq2.constrained.simml}) 
reduces to the modified covariate model (\ref{the.MC}) when using the framework  (\ref{LS4}) for the optimization of the dimension reduction vector $\bm{\beta}$. 
The approximation model (\ref{eq2.constrained.simml})  is a special case of 
the dimension reduction model (\ref{CSIM.model})  with the linear $a$-specific functions 
 $g_{0a}(u) = \gamma_a u$ $(a \in \mathcal{A})$. 
 Therefore, the modified covariate method can be viewed as a special case of the approach that estimates 
a vector in $S_{\mathcal{C}|\bm{X}}$, 
when we restrict the $a$-specific functions $g_a$ to be linear, 
 and restrict the case to $L =2$.

\section{A SEMIPARAMETRIC MODEL} \label{sec.semipar.model}

A semiparametric generalization of  the  linear  rank-1 approximation model (\ref{eq2.constrained.simml}) to model (\ref{the.decomposition})
can be defined based on replacing  the set of $a$-specific slopes $\{\gamma_a \in \mathbb{R} \}_{a \in \mathcal{A}}$ 
 in model (\ref{eq2.constrained.simml}) to a set of 
nonpametrically-defined $a$-specific
 functions $\{ g_a  \in \mathcal{H}^{(\bm{\beta})}  \}_{a \in \mathcal{A}}$.  
 Note, for each fixed $\bm{B} \in \Theta_q$, 
 the condition $  \mathbb{E}[ g_A(\bm{B}^\top \bm{X}) | \bm{B}^\top\bm{X} ] =0$  
 implies $\mathbb{E}[ g_A(\bm{B}^\top \bm{X}) | \bm{X}]   = \mathbb{E}[  \mathbb{E}[ g_A(\bm{B}^\top \bm{X}) | \bm{B}^\top\bm{X} ] | \bm{X}]  = 0$ 
 by an application of the iterated expectation rule to condition on $\bm{B}^\top\bm{X}$, 
which in turn implies the constraint in (\ref{LS4}). 
Then, with $\bm{B} = \bm{\beta} \in \Theta_1$, 
 the optimization  (\ref{LS4}) can be simplified to: 
 \begin{equation} \label{the.condition2}
\begin{aligned}
& 
\underset{g_a \in \mathcal{H}^{(\bm{\beta})}, \bm{\beta} \in \Theta_1 }{\text{argmin}}
& &\mathbb{E} \big[ \big(Y -  g_{A}( \bm{\beta}^\top \bm{X}) \big)^2 \big]  \\
& \text{subject to} & & \mathbb{E}[ g_A(\bm{\beta}^\top \bm{X}) |\bm{\beta}^\top\bm{X} ] =0. 
\end{aligned}
\end{equation}

 Under model (\ref{CSIM.model}), solving  (\ref{the.condition2}) yields a vector, say  $\bm{\beta}_0  \in \Theta_1$,
  that approximates a vector in  $S_{\mathcal{C}|\bm{X}} = \mbox{span}(\bm{B}_0)$. 
 If  $q=1$, then 
 $\bm{\beta}_0 = \bm{B}_0$, and  
if $q >1$,  then $\mbox{span}(\bm{\beta}_0)$
is the best  rank-1 approximation to the $\mbox{span}(\bm{B}_0)$ of the interaction term of model (\ref{CSIM.model}) in $L^2$: 
%
 \begin{equation}\label{the.criterion4}
\begin{aligned}
& \underset{g_a \in \mathcal{H}^{(\bm{\beta})}, \bm{\beta} \in \Theta_1 }{\text{argmin}}   \mathbb{E}\left[ Y^2   - 2 g_A(\bm{\beta}^\top \bm{X}) \big( \mu(\bm{X})  + g_{0A}(\bm{B}_0^\top \bm{X})\big)  +  \big(g_A(\bm{\beta}^\top \bm{X}) \big)^2 \right]  \\
=& \underset{g_a \in \mathcal{H}^{(\bm{\beta})}, \bm{\beta} \in \Theta_1 }{\text{argmin}}  \mathbb{E}\left[ Y^2  - 2 g_A(\bm{\beta}^\top \bm{X})  g_{0A}(\bm{B}_0^\top \bm{X}) +  \big( g_A(\bm{\beta}^\top \bm{X}) \big)^2     \right]  \\
=& \underset{g_a \in \mathcal{H}^{(\bm{\beta})}, \bm{\beta} \in \Theta_1 }{\text{argmin}}   \mathbb{E}\left[  \big( g_A(\bm{\beta}^\top \bm{X}) - g_{0A}(\bm{B}_0^\top \bm{X}) \big)^2      \right], 
\end{aligned}
\end{equation}  
where the first line comes from expanding the squared error criterion in (\ref{the.condition2}) and the assumed model (\ref{CSIM.model}), 
and the second line follows from an application of the iterated expectation rule to condition on $\bm{X}$ and that $\mu(\bm{X}) \mathbb{E}[ g_A(\bm{\beta}^\top \bm{X}) | \bm{X}]  = 0$, implied by the constraint on $\{g_a\}_{a \in \mathcal{A}}$  
 in  (\ref{the.condition2}). 

The appealing feature of  the optimization framework (\ref{LS4}) 
is that the 
term $\mu(\bm{X})$ in  (\ref{CSIM.model}) 
 does not have to be specified when 
approximating a vector in  $S_{\mathcal{C}|\bm{X}}$. 
Due to the nonlinearity of the functions $\{g_a\}_{a \in \mathcal{A}}$, 
a closed form solution of (\ref{the.condition2}) 
 is not available. 
We briefly sketch below a procedure for solving  (\ref{the.condition2}). 

Generally, we can employ an iterative procedure that alternates  between: 
1)  (given a $\bm{\beta}$,) solving a penalized least square regression  with a basis expansion 
for each of the functions $\{g_a  \in \mathcal{H}^{(\bm{\beta})} \}_{a \in \mathcal{A}}$,   
with an appropriate penalization for the function smoothness;    
and 2) (given $\{g_a\}_{a \in \mathcal{A}}$,) estimating  $\bm{\beta} \in \Theta_1$. 
The constraint on the functions $\{g_a\}_{a \in \mathcal{A}}$   in (\ref{the.condition2}) 
can be absorbed into their basis construction 
through reparametrization, as we describe next. 
Suppose we are given $(\bm{\beta}^\top \bm{x}_i, a_i)$ $(i=1,\ldots,n)$ for a fixed $\bm{\beta}$. 
We can represent  $g_{a_i}(\bm{\beta}^\top \bm{x}_i)$ $(i=1,\ldots,n)$  
based on a $d$-dimensional basis 
$\Psi(\cdot) \in \mathbb{R}^{d}$ (e.g., $B$-spline basis on evenly spaced knots on a bounded domain): 
\begin{equation} \label{eq.f}
g_{a_i}(\bm{\beta}^\top \bm{x}_i) = \Psi(\bm{\beta}^\top \bm{x}_i)^\top \bm{\theta}_{a_i}   \quad (i=1,\ldots,n)
\end{equation}
for a set of 
unknown basis coefficients $\{\bm{\theta}_{a}  \in \mathbb{R}^d \}_{a \in \mathcal{A}}$. 
We impose 
 the following restriction  on (\ref{eq.f}) to  satisfy the required constraint in  (\ref{the.condition2}), 
\begin{equation}\label{beta.t.condition}
\sum_{a=1}^L \pi_a \bm{\theta}_{a} =  \bm{\pi} \bm{\theta}  = \bm{0}.
\end{equation}
Here  $\bm{\theta} := (\bm{\theta}_{1}^\top, \ldots, \bm{\theta}_{L}^\top )^\top \in \mathbb{R}^{dL}$ is the vectorized version of the basis coefficients 
$\{\bm{\theta}_{a}\}_{a \in \mathcal{A}}$ in (\ref{eq.f}), 
the matrix $\bm{\pi} := \left(\pi_1 \bm{I}_{d};  \ldots; \pi_L \bm{I}_{d}\right)$ 
is the $d \times dL$ constraint matrix associated with the coefficient $\bm{\theta}$, in which $\bm{I}_{d}$ denotes the $d \times d$ identity matrix, 
and $\bm{0} $ is the length-$d$ vector of zeros. 
Condition (\ref{beta.t.condition}) indicates $\mathbb{E}[\bm{\theta}_A] = \bm{0}$, and is a sufficient condition for any set of the functions 
of the form (\ref{eq.f}) to satisfy the constraint in (\ref{the.condition2}).

Let the $n \times d$  matrix   $\bm{D}_{a} $ $(a \in \mathcal{A})$ 
denote the (treatment $a$-specific) evaluation matrix of the basis $\Psi(\cdot)$ on $\bm{\beta}^\top \bm{x}_i$ $(i=1,\ldots,n)$, 
in which $i$th row  is  the $1 \times d$ vector $\Psi(\bm{\beta}^\top \bm{x}_i)^\top$  if  $a_i = a$, 
and a row of zeros $\bm{0}^\top$ if $a_i \ne a$. 
Then, the column-wise concatenation 
of the design matrices $\{ \bm{D}_{a}\}_{a \in \mathcal{A}}$, 
i.e., the $n \times dL $ matrix
$\bm{D}  = (\bm{D}_{1}; \ldots; \bm{D}_{L})$, 
defines the model matrix associated with the  model coefficient  
$\bm{\theta} \in \mathbb{R}^{dL}$, vectorized across 
$\{\bm{\theta}_{a}\}_{a \in \mathcal{A}}$ in (\ref{eq.f}).

To define a penalty associated with $\bm{\theta} \in \mathbb{R}^{dL}$, 
we write 
$\bm{S}_a = (\bm{\delta}_a^\top \otimes \bm{P})^\top (\bm{\delta}_a^\top \otimes \bm{P})$ $(a \in \mathcal{A})$, 
where 
$\bm{P}$ represents 
a ``square root'' of some penalty matrix associated with each $\bm{\theta}_a \in \mathbb{R}^d$ 
(e.g.,  a   second order $P$-splines difference penalty 
 \citep{Eilers1996} of dimension $(d-2) \times d$), 
 the vector 
 $\bm{\delta}_a \in \mathbb{R}^L$  
is the vector of all zeros 
except its $a$th element equal to $1$, and 
  $\otimes$ represents the Kronecker product.

Given a set of tuning parameters $\{ \lambda_a \ge 0 \}_{a \in \mathcal{A}}$, 
an empirical  criterion function associated with the constrained optimization problem (\ref{the.condition2}) can be written as: 
\begin{equation} \label{Q.criterion}
Q(\bm{\theta}, \bm{\beta}) \ 
=
\ \lVert  \bm{Y}_{n \times 1}  - \bm{D} \bm{\theta} \rVert^2 \ + 
\
 \sum_{a=1}^L  \lambda_a  \bm{\theta}^\top \bm{S}_a \bm{\theta}, 
\end{equation} 
constrained by 
(\ref{beta.t.condition}) and $\bm{\beta} \in \Theta_1$. 
The linear constraint  (\ref{beta.t.condition}) $\bm{\pi} \bm{\theta} = \bm{0}$  can be  absorbed into the model matrix $\bm{D}$ 
and the penalty matrices $\bm{S}_a$ $(a = 1,\ldots,L)$ as follows. 
We can create a $dL \times d(L-1)$ (orthonormal) basis matrix $\bm{Z}$, such that if set $\bm{\theta} = \bm{Z} \tilde{\bm{\theta}}$  for any (unconstrained) vector $\tilde{\bm{\theta}} \in \mathbb{R}^{d(L-1)}$, then 
$\bm{\pi} \bm{\theta} = \bm{0}$,  thus satisfying (\ref{beta.t.condition}). 
Such a basis matrix 
$\bm{Z}$ 
can be found by a QR decomposition of 
 $\bm{\pi}^\top$. 
Given such a basis $\bm{Z}$ of the null space of (\ref{beta.t.condition}), 
we can reparametrize 
  (\ref{Q.criterion}) 
with respect to the unconstrained vector $\tilde{\bm{\theta}}$ (and $\bm{\beta} \in \Theta_1$),  
 by setting  $\tilde{\bm{D}} \leftarrow  \bm{D}  \bm{Z}$ 
and $\tilde{\bm{S}}_a \leftarrow \bm{Z}^\top \bm{S}_a \bm{Z}$ $(a = 1,\ldots, L)$, 
which yields: 
\begin{equation} \label{Q.criterion2}
\begin{aligned}
\tilde{Q}(\tilde{\bm{\theta}}, \bm{\beta}) \ 
&=
\ \lVert  \bm{Y}_{n \times 1}  - \tilde{\bm{D}} \tilde{\bm{\theta}} \rVert^2 + 
\sum_{a=1}^L  \lambda_a  \tilde{\bm{\theta}}^{\top}  \tilde{\bm{S}}_a  \tilde{\bm{\theta}}, 
\end{aligned}
\end{equation} 
where $\tilde{\bm{\theta}} \in \mathbb{R}^{d(L-1)}$ and $\bm{\beta} \in \Theta_1$. 
The smoothing parameters $\lambda_a$ in (\ref{Q.criterion2}) 
 can be optimized, for example, via restricted maximum likelihood (REML) estimation, 
 and the associated profile minimizer  $\hat{\tilde{\bm{\theta}}}  \in \mathbb{R}^{d(L-1)}$  
 of (\ref{Q.criterion2}) given a fixed $\bm{\beta}$ results in a set of 
 estimates  $\hat{g}_a(\cdot) = \Psi(\cdot)^\top \hat{\bm{\theta}}_{a}$ $(a=1,\ldots,L)$ for the $a$-specific smooths $g_a(\cdot)$ $(a=1,\ldots,L)$ in (\ref{eq.f}), 
where $(\hat{\bm{\theta}}_1^\top, \ldots, \hat{\bm{\theta}}_L^\top)^\top   := \bm{Z} \hat{\tilde{\bm{\theta}}}$. 
To optimize (\ref{Q.criterion2}) over $\bm{\beta} \in \Theta_1$ given $\hat{g}_a(\cdot)$ $(a=1,\ldots,L)$, 
we can perform a linear approximation 
of $\hat{g}_{a_i}( \bm{\beta}^\top \bm{x}_i )$ with respect to $\bm{\beta}$ 
at the current ($k$th) iterate, say  
$\bm{\beta}^{(k)} \in \Theta_1$, 
and approximate the squared error part of (\ref{Q.criterion2}): 
\begin{equation} \label{Q.hat2}
\begin{aligned}
\lVert  \bm{Y}_{n \times 1}  - \tilde{\bm{D}} \tilde{\bm{\theta}} \rVert^2 \
& \approx  \    \sum_{i=1}^n \left(y_i -   \hat{g}_{a_i}( \bm{\beta}^{(k)\top}\bm{x}_i )   -   \dot{\hat{g}}_{a_i}(\bm{\beta}^{(k)\top}\bm{x}_i )  (\bm{\beta} - \bm{\beta}^{(k)})^\top \bm{x}_i
\right)^2, 
\end{aligned}
\end{equation}
where $\dot{\hat{g}}_{a}(\cdot)$ denotes the first derivative of $\hat{g}_{a}(\cdot)$. 
The right-hand side of 
 (\ref{Q.hat2}) can be minimized over $\bm{\beta} \in \mathbb{R}^p$ via a  least squares regression, and 
 the associated minimizer $\bm{\beta}^{(k+1)} \in \mathbb{R}^p$ is then scaled to satisfy $\bm{\beta}^{(k+1)} \in \Theta_1$.  
 We can iterate between optimizing 
 $\tilde{\bm{\theta}} \in \mathbb{R}^{d(L-1)}$ and $\bm{\beta} \in \Theta_1$, 
 until $\lVert (\bm{\beta}^{(k+1)}  - \bm{\beta}^{(k)})/\bm{\beta}^{(k+1)} \rVert$   
 is less than a pre-specified convergence tolerance. 
 
The consistency, the details of the estimation procedure and extensive numerical examples for this semiparametric approach are given in \cite{CSIM}.

\section{GEOMETRIC INTUITION} \label{sec.intuition}

In this section, we will provide some geometric intuition  behind 
the optimization approach (\ref{the.condition2}) to approximating  the interaction effect term of model (\ref{the.decomposition}).  
It is straightforward to verify that, for each fixed $\bm{\beta} \in \Theta_1$, 
the minimizer  $\{ g_a^\ast \}_{a \in \mathcal{A}}$ 
of (\ref{the.condition2}) satisfies: 
\begin{equation}\label{g.solution}
 g_{a}^\ast( {\bm{\beta}}^\top \bm{X})  =  \mathbb{E}[Y |  {\bm{\beta}}^\top \bm{X}, A=a] -  \mathbb{E}[Y |  {\bm{\beta}}^\top \bm{X}]  \quad (a \in \mathcal{A}). 
\end{equation} 
The first term 
$\mathbb{E}[Y |  {\bm{\beta}}^\top \bm{X}, A=a]$ $(a \in \mathcal{A})$  in  (\ref{g.solution}) 
is the treatment $a$-specific $L^2$ projection of $Y$ onto $\mathcal{H}^{(\bm{\beta})}$,  
and the second term 
$-  \mathbb{E}[Y |  {\bm{\beta}}^\top \bm{X}] $ 
in (\ref{g.solution}) 
``shifts'' this unconstrained treatment $a$-specific $L^2$ projection 
to satisfy the constraint  in (\ref{the.condition2}). 
This results in orthogonality,
$g_{A}^\ast( {\bm{\beta}}^\top \bm{X}) \perp \mu(\bm{X})$, 
 between (\ref{g.solution}) and the unspecified term $\mu(\bm{X})$ in (\ref{the.decomposition}), for each fixed $\bm{\beta}$. 

\begin{figure}[h] 
\begin{center} 
\includegraphics[width=3.4in, height = 2.1in]{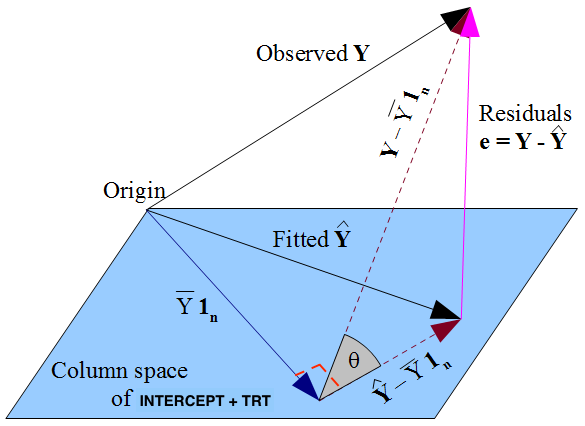} 
\end{center}
\caption[]{
In the regression $\mathbb{E}[Y | A ]$ of  $Y$ on  $A$ (the treatment variable), the fitted $\hat{\bm{Y}}$  is the orthogonal projection of the observed $\bm{Y}$ onto   the column space spanned by  the treatment $A$ and the intercept ``1'' which is represented by the blue plane. The fitted vector for the ``1'' (i.e., the intercept)-only model $\mathbb{E}[Y]$ is  represented by $\bar{Y} \bm{1}_n$. The magnitude of the interaction effect between $A$ and ``1''  
 is quantified by the squared length of the vector $ \hat{\bm{Y}} - \bar{Y} \bm{1}_n$. 
 } \label{fig.geometric.intuition}
\end{figure}

For illustration, 
we consider a very simple example 
 of  regressing $Y$ on 
 the treatment variable $A \in \mathcal{A}$ with no covariate $\bm{X}$ (i.e., the $\bm{X}$ corresponds to the intercept ``1''). 
In this simple setting,  
the solution $\{ g_a^\ast \}_{a \in \mathcal{A}}$ 
 in (\ref{g.solution}) are just constants: 
\begin{equation}\label{simple.proj}
g_{a}^\ast = \mathbb{E}[Y | A=a] - \mathbb{E}[Y] \quad (a \in \mathcal{A}). 
\end{equation}

Given sample data $(y_i, a_i)$ $(i=1,\ldots,n)$, 
let $\bm{Y} = (y_1,\ldots, y_n)^\top$ denote the (length-$n$) observed vector of  responses. 
The second term  $\mathbb{E}[Y]$ 
on the right-hand side  of  
(\ref{simple.proj}) is represented by the (length-$n$) vector
$ \bar{Y} \bm{1}_n$, in which 
$\bar{Y} = \sum_{i=1}^n y_i/n$  is the grand mean of $Y$, and 
$\bm{1}_n = (1, 1, \ldots,1 )^\top$. 
The first term $\mathbb{E}[Y |  A=a]$ $(a \in \mathcal{A})$ 
on the right-hand side  of   (\ref{simple.proj}) is represented by 
the (length-$n$) vector 
$\hat{\bm{Y}} = (\hat{Y}_1,\ldots, \hat{Y}_n)^\top$, 
where $\hat{Y}_i = \sum_{a=1}^L  1_{(a_i=a)} \bar{Y}^{(a)}$ $(i=1,\ldots,n)$, 
with 
$\bar{Y}^{(a)} = \sum_{i=1}^n y_i 1_{(a_i=a)}/\sum_{i=1}^n 1_{(a_i=a)}$ 
denoting the treatment $a$-specific mean. 
The fitted function $g_A^\ast$  
in (\ref{simple.proj}) is 
 thus represented (in $\mathbb{R}^n$) by the vector
$\hat{\bm{Y}} - \bar{Y} \bm{1}_n$. 
These three vectors
($ \bar{Y}\bm{1}_n$, $\hat{\bm{Y}}$ and $\hat{\bm{Y}} - \bar{Y} \bm{1}_n$) 
 in $\mathbb{R}^n$ are represented in  Figure~\ref{fig.geometric.intuition}. 
 
By constraint in the second line of (\ref{the.condition2}), $\mathbb{E}[g_{A}^\ast] =0$ 
and thus, $\mbox{var}[  g_{A}^\ast] = \mathbb{E}[ (g_{A}^\ast)^2]$ 
which is represented by $ \lVert \hat{\bm{Y}} - \bar{Y} \bm{1}_n \rVert^2$ in Figure~\ref{fig.geometric.intuition}.    
Notice that 
the fitted vector, $\hat{\bm{Y}} - \bar{Y} \bm{1}_n$, 
is orthogonal to the ``nuisance'' vector $\bar{Y} \bm{1}_n$ in Figure~\ref{fig.geometric.intuition}.

Intuitively, the ``effect'' of intercept ``1'' in the intercept-only model is to average  the response $\bm{Y}$, 
which results in the fit $\bar{Y} \bm{1}_n$ in Figure~\ref{fig.geometric.intuition}. 
The variance $\lVert \hat{\bm{Y}} - \bar{Y} \bm{1}_n \rVert^2  \sim  \mbox{var}[  g_{A}^\ast]$, 
where $\hat{\bm{Y}}$ is the vector of treatment $a$-specific averages,  
quantifies  the magnitude of 
how much the ``effect'' of intercept ``1'' (i.e., the grand averaging) 
 is modified by the variable $A$, 
 and hence the variance
  $\lVert \hat{\bm{Y}} - \bar{Y} \bm{1}_n \rVert^2$  quantifies 
the intensity of the ``interaction effect'' between the intercept ``1'' and  $A$.  
Analogously, in the optimization framework  (\ref{the.condition2}), 
given a candidate $\bm{\beta} \in \Theta_1$,   
the variance of the profile minimizer $\{ g_a^\ast \}_{a \in \mathcal{A}}$ 
in (\ref{g.solution}), i.e., 
$\mbox{var}\left[ g_{A}^\ast({\bm{\beta}}^\top \bm{X}) \right] = \mathbb{E}\left[ \{g_{A}^\ast({\bm{\beta}}^\top \bm{X})\}^2 \right]$, 
quantifies  the magnitude of the interaction effect between the candidate linear predictor (i.e., single-index) ${\bm{\beta}}^\top \bm{X}$ 
and the variable $A$. 
This variance of the interaction effect  is to be maximized over $\bm{\beta} \in \Theta_1$, 
as in the case of maximizing  the variance (\ref{the.numerator.method}). 
 
When ${\bm{\beta}}^\top \bm{X}$ replaces the intercept ``1'', 
for each $\bm{\beta} \in \Theta_1$, 
the blue plane  in  Figure~\ref{fig.geometric.intuition}, 
represents the Hilbert space 
of measurable functions 
of $({\bm{\beta}}^\top \bm{X}, A$).  
Maximizing the variance of the $\bm{X}$-by-$A$ interaction effect, i.e., 
$\mbox{var}\left[ g_{A}^\ast({\bm{\beta}}^\top \bm{X}) \right]$,   
over ${\bm{\beta}} \in \Theta_1$ corresponds to adjusting  
 the blue plane of Figure~\ref{fig.geometric.intuition},  
in such a way that the blue plane 
minimizes the angle $\theta$ formed by the hypotenuse $\bm{Y} - \bar{Y} \bm{1}_n$ and the adjacent $\hat{\bm{Y}} - \bar{Y} \bm{1}_n$ (i.e., the $\theta$ formed by the two dashed lines in  Figure~\ref{fig.geometric.intuition}). 
Or equivalently,  it corresponds to maximizing the cosine of the angle $\theta$ (over $\bm{\beta} \in \Theta_1$), thereby maximizing the length of the vector $\lVert \hat{\bm{Y}} - \bar{Y} \bm{1}_n \rVert^2$.

Finally, we note that  the two centered vectors 
$\hat{\bm{Y}} - \bar{Y} \bm{1}_n$ 
and 
$\bm{Y} - \bar{Y} \bm{1}_n$ 
  (i.e., the two dashed lines in Figure~\ref{fig.geometric.intuition}) 
  correspond to 
  the fitted ($\hat{\bm{Y}}$) and the observed ($\bm{Y}$) vectors, respectively, 
 centered by the intercept vector $(\bar{Y} \bm{1}_n)$. 
Without centering by the intercept $\bar{Y} \bm{1}_n$, there is no 
 Pythagorean-type sum of squares decomposition:  
 \begin{equation} \label{partition2}
\lVert   \bm{Y}  -  \bar{Y} \bm{1}_n \rVert^2 
=   \lVert \bm{Y} - \hat{\bm{Y}} \rVert^2  + \lVert \hat{\bm{Y}} - \bar{Y} \bm{1}_n  \rVert^2,   
\end{equation}
in which the second term, 
$ \lVert \hat{\bm{Y}} - \bar{Y} \bm{1}_n \rVert^2$, quantifies 
the $A$-by-``$1$'' interaction effect. 
Analogously,  the ``shifting'' component $-\mathbb{E}[Y | {\bm{\beta}}^\top \bm{X}]$  in (\ref{g.solution}) plays the role of an   
``intercept.'' 
Centering by the function $\mathbb{E}[Y | {\bm{\beta}}^\top \bm{X}]$
allows the following Pythagorean-type  decomposition and isolates the variance associated with the interaction effect: 
 \begin{equation} \label{partition3}
\mathbb{E}\big[ \big(  Y  - \mathbb{E}[Y | {\bm{\beta}}^\top \bm{X}] \big)^2 \big] = 
 \mathbb{E}\big[ \big( Y  - \mathbb{E}[Y | {\bm{\beta}}^\top \bm{X}, A]\big)^2 \big] + 
 \mathbb{E}\big[ \big( \mathbb{E}[Y | {\bm{\beta}}^\top \bm{X},A]  - \mathbb{E}[Y | {\bm{\beta}}^\top \bm{X}] \big)^2 \big], 
\end{equation}
where the magnitude of 
 the $A$-by-${\bm{\beta}}^\top \bm{X}$ interaction effect is quantified  by the second term   
 $\mathbb{E}\big[\big(g_A^\ast({\bm{\beta}}^\top \bm{X})\big)^2 \big] =  \mathbb{E}\big[ \big( \mathbb{E}[Y | {\bm{\beta}}^\top \bm{X},A]   - \mathbb{E}[Y | {\bm{\beta}}^\top \bm{X}] \big)^2 \big]$ (see (\ref{g.solution}), for this equality), 
that is to be maximized over ${\bm{\beta}} \in \Theta_1$.

\section{SUFFICIENT REDUCTION FOR INTERACTIONS BETWEEN COVARIATES AND A CONTINUOUS VARIABLE}
\label{continuous.treatment}

 In this section, we extend the semiparametric dimension reduction model to the case where the variable $A$ is defined on a compact continuum $\mathcal{A}\subset \mathbb{R}$. 
 In this case, 
we can define  a {\em contrast function} $\bm{c}  \in \{ \bm{c} :  0 < \int_{\mathcal{A}}{\bm{c}}^2(a) da,  \int_{\mathcal{A}}{\bm{c}}(a) da =0  \} \subset L^2(\mathcal{A})$, and the associated mean contrast function $\mathcal{C}(\bm{X};  {\bm c}) := \int_{\mathcal{A}} \bm{c}(a) \mathbb{E}\left[ Y |  \bm{X} , A= a \right] da$. 
We consider the dimension reduction model of form 
  \begin{equation} \label{SIMSL.model}
\mathbb{E}\left[Y \mid \bm{X}, A\right] 
 =  \mu(\bm{X})  
 +  
  g_{0}\big( \bm{B}_0^\top \bm{X}, A \big),
\end{equation}
where 
the smooth function $g_0$ is a ($q+1$ dimensional)  function of $( \bm{B}_0^\top \bm{X}, A)$, 
with the resulting function 
$g_{0}( \bm{B}_0^\top \bm{X}, A)$
  determining the $\bm{X}$-by-$A$ interaction effect;  the term $\mu(\bm{X})$ 
  represents an unspecified main effect of $\bm{X}$. 
As in (\ref{CSIM.model}), we assume   
$\mathbb{E}[ g_{0}\big( \bm{B}_0^\top \bm{X}, A \big) | \bm{X}] = 0$ 
and $\bm{B}_0 \in \Theta_q$, 
for model identifiability. 
Let  $ \mathcal{H}_{q+1}^{(\bm{B})}$ 
denote the Hilbert space of measurable functions of $(\bm{B}^\top \bm{X}, A)$  
given each $\bm{B}  \in \Theta_q$, 
and in model (\ref{SIMSL.model}), we assume $g_0 \in  \mathcal{H}_{q+1}^{(\bm{B}_0)}$.

As in Section~\ref{sec.semipar.model}, 
we focus on a rank-1 (i.e., $q=1$) approximation model with a vector $\bm{\beta} \in \Theta_1$. 
For a continuous $A \in \mathcal{A}$, 
we modify the optimization framework (\ref{the.condition2}) that utilizes a set of treatment $a$-specific 1-D smooths $\{ g_a \in \mathcal{H}^{(\bm{\beta})}\}_{a \in \mathcal{A}}$  
to that with a  single $2$-D smooth $g \in \mathcal{H}_2^{(\bm{\beta})}$: 
\begin{equation} \label{sub.problem2}
\begin{aligned}
(\hat{g}, \hat{\beta}) \quad = \quad  & \underset{g \in \mathcal{H}_{2}^{(\bm{\beta})}, \bm{\beta} \in \Theta_1}{\text{argmin}}
& &E \big[ \big(Y - g\big(\bm{\beta}^{\top}\bm{X}, A )\big)^2 \big]  \\
& \text{subject to} & &\mathbb{E}\left[g(\bm{\beta}^\top \bm{X}, A) |\bm{\beta}^\top \bm{X} \right] = 0 
\end{aligned}
\end{equation}
We briefly sketch below an iterative procedure to solve  (\ref{sub.problem2}). 
As in Section~\ref{sec.semipar.model}, 
we can alternate between estimation of $\bm{\beta} \in \Theta_1$ and $g \in \mathcal{H}_2^{(\bm{\beta})}$.  
Again, the constraint in (\ref{sub.problem2}) can be  absorbed into a (tensor-product) basis  representation of the 2-D smooth $g$ by reparametrization.

Denoting $\eta = \bm{\beta}^\top \bm{X}$, 
 (although other linear smoothers can also be utilized),
let us focus on tensor products of $B$-splines \citep{deBoor} 
to represent the 
smooth $g(\eta, A)$ 
 in (\ref{sub.problem2}) for each fixed $\bm{\beta}$, with  a set of separate difference penalties applied to the coefficients of the basis along the $\eta$ and $A$ axes, i.e., the tensor-product P-splines \citep[][]{PSR.interaction}. 
We shall use the tensor product of $2$ univariate cubic $B$-splines, 
say $\Psi$ and $\check{\Psi}$, 
with $d$ and $\check{d}$  equally-spaced $B$-spline basis functions placed along the $\eta$ and $A$ axes respectively. 
Associated with the $d$ and $\check{d}$-dimensional marginal  bases are $d \times d$ and $\check{d}  \times \check{d}$ roughness penalty matrices, which we write as $S$ and $\check{S}$ respectively.
 
For fixed $\eta_i = \bm{\beta}^\top \bm{x}_i$ $(i=1,\ldots,n)$,  
let us write the $n \times d$ (and $n \times \check{d}$) $B$-spline  design matrix 
$\bm{\Psi}$ (and $\check{\bm{\Psi}}$), 
in which its $i$th row  is $\bm{\Psi}_i = \Psi(\eta_i)^\top$ (and $\check{\bm{\Psi}}_i = \check{\Psi}(a_i)^\top$). 
Then, for each fixed $\bm{\beta}$, a flexible surface $g$ in (\ref{sub.problem2}) can be approximated at the points $(\eta_i, a_i)$ $(i=1,\ldots, n)$ \citep{VCSISR}, 
\begin{equation} \label{tensor.product}
\mbox{vec}\big\{  g(\eta_i, a_i)  \big\} =  g( \eta_{n \times 1}, a_{n \times 1})  = \bm{D} \bm{\theta},
\end{equation}
with the  $n \times d\check{d}$ tensor product model matrix 
\begin{equation} \label{eq.D}
\bm{D} =  \left( \bm{\Psi} \otimes  \bm{1}_{\check{d}}^\top \right) \odot \left(  \bm{1}_{d}^\top  \otimes  \check{\bm{\Psi}}  \right), 
\end{equation}
where $\odot$ denotes element-wise multiplication of matrices 
and $\bm{\theta} \in \mathbb{R}^{d \check{d}}$ is an unknown coefficient vector associated with  the function $g$.

\cite{Wood2006} noted that constructing tensor products of the form (\ref{tensor.product}) 
is a general approach to producing tensor product smooths of several variables, constructed from 
the univariate (marginal) bases  $\bm{\Psi}$ and $\check{\bm{\Psi}}$ separately, 
and can be utilized for a general $q> 2$ case. 
Similarly, the roughness penalty matrices associated with 
the tensor product model (\ref{tensor.product})
can be constructed from 
the individual roughness penalty matrices, $S$ and $\check{S}$, 
and are given by 
$\bm{S} = S \otimes  \bm{I}_{\check{d}}$ and $\check{\bm{S}} = \bm{I}_{d} \otimes  \check{S}$, 
for the axis directions $\eta$ and $A$, respectively; here, 
$\bm{I}$ denotes the identity matrix, and both $\bm{S}$ and  $\check{\bm{S}}$
are square matrices of dimension $d \check{d}$.

We now impose  the constraint in  (\ref{sub.problem2}) on  the smooth 
$g$ under the tensor product representation  (\ref{tensor.product}). 
For each fixed $\bm{\beta}$, 
the constraint in (\ref{sub.problem2}) on $g$ amounts to 
excluding the main effect of $\eta = \bm{\beta}^\top \bm{X}$
from the smooth $g$. 
We deal with this by reparametrizing the representation (\ref{tensor.product}).  
Consider the following sum-to-zero (over the observed values) constraint for the marginal basis 
of $A$: 
\begin{equation}\label{lin.constr} 
\bm{1}^\top \check{\bm{\Psi}} \check{\bm{\gamma}} = 0,
\end{equation} 
for any $ \check{\bm{\gamma}} \in \mathbb{R}^{\check{d}}$, 
where $\bm{1}$ is a length-$n$ vector of 1's. 
With the constraint (\ref{lin.constr}), the linear smoother associated with the basis  $\check{\bm{\Psi}}$ cannot reproduce constant functions \citep{GAM}.  
That is, 
the linear constraint (\ref{lin.constr}) removes the span of constant functions  
 from the span of the marginal basis $\check{\bm{\Psi}}$, with the result that 
the tensor product basis, 
 $\bm{D}$ in (\ref{tensor.product}),  
  will not include the main effect of $\eta$  
   that results from the product of the marginal basis $\bm{\Psi}$ (associated with $\eta$)
  with the constant  function in the span of  
  the other marginal basis $\check{\bm{\Psi}}$ (associated with $A$). 
Therefore, the resultant fit of the 2-D smooth $g$, under representation (\ref{tensor.product}) subject to (\ref{lin.constr}), 
excludes the main effect of $\eta$. 
  See Section 5.6 of \cite{Wood2017} for additional details. 
Incorporating such a linear constraint (\ref{lin.constr}) on the model matrix $\bm{D}$ in (\ref{tensor.product})  is given below. 

The key is to find an (orthogonal) 
basis  for the null space of the constraint  (\ref{lin.constr}), 
and then 
absorb the constraint into construction of $\bm{D}$ in (\ref{eq.D}). 
To be specific, 
we can create a $\check{d} \times (\check{d}-1)$ matrix $\bm{Z}$, such that if  $\check{\bm{\gamma}} = \bm{Z} \check{\bm{\gamma}}^\ast$  for any $\check{\bm{\gamma}}^\ast \in \mathbb{R}^{\check{d}-1}$, then 
$\bm{1}^\top \check{\bm{\Psi}} \check{\bm{\gamma}} = 0$,  satisfying the constraint (\ref{lin.constr}). 
Such a  matrix 
$\bm{Z}$ can be found by a QR decomposition of $\check{\bm{\Psi}}^\top \bm{1}$. 
Then, we can 
reparametrize  the marginal basis
 $\check{\bm{\Psi}}$ by 
$\check{\bm{\Psi}}^\ast \leftarrow \check{\bm{\Psi}} \bm{Z}$ 
(and the associated penalty matrix by $\check{S}^\ast \leftarrow \bm{Z}^\top \check{S} \bm{Z}$) and absorb the constraint (\ref{lin.constr}) into its basis construction.
Accordingly, the resulting reparametrized model matrix (\ref{eq.D}) is given by 
$\bm{D}^\ast \leftarrow  \left( \bm{\Psi} \otimes  \bm{1}_{\check{d}-1}^\top \right) \odot \left(  \bm{1}_{d}^\top  \otimes  \check{\bm{\Psi}}^\ast  \right)$ 
and the associated penalty matrices are  
$\bm{S}^\ast \leftarrow S \otimes  \bm{I}_{\check{d}-1}$ and $\check{\bm{S}}^\ast \leftarrow \bm{I}_{d} \otimes  \check{S}^\ast$, 
for the axis directions $\eta$ and $A$, respectively; 
$\bm{\theta} \in \mathbb{R}^{d\check{d}}$ in (\ref{tensor.product})  is also reparametrized to $\bm{\theta}^\ast \in \mathbb{R}^{d (\check{d}-1)}$.  

This sum-to-zero reparametrization enforcing  (\ref{lin.constr}) to representation (\ref{tensor.product})  is simple, and creates a term $\mbox{vec}\big\{  g(\eta_i, a_i)  \big\} =  \bm{D}^\ast \bm{\theta}^\ast$ 
 that specifies such pure $\bm{X}$-by-$A$ interactions (plus the $A$ main effect) that are orthogonal to the $\bm{X}$ main effect.  
Provided that the orthogonality constraint issue is addressed, for  each fixed $\bm{\beta}$,   
the criterion  (\ref{sub.problem2})  can be represented by a penalized least squares criterion, 
$Q(\bm{\theta}^\ast, \bm{\beta}) =
 \lVert  Y_{n \times 1}  - \bm{D}^\ast \bm{\theta}^\ast \rVert^2 +  \lambda \bm{\theta}^{\ast\top} \bm{S}^\ast \bm{\theta}^\ast 
 + \check{\lambda}  \bm{\theta}^{\ast\top} \check{\bm{S}}^\ast \bm{\theta}^\ast$, 
in which the smoothing  parameters $\lambda$ and $\check{\lambda}$ can be estimated by,  for example, REML.
Similar to Section~\ref{sec.semipar.model}, 
we can iterate between optimizing $\bm{\theta}^\ast$ and $\bm{\beta}$ until convergence.

 
 \section{MULTIPLE PROJECTIONS FOR SUFFICIENT REDUCTION }\label{multiple.projection}

 We have so far focused on single-dimensional approximations (i.e., $q=1$ case). 
In this section, we consider generalizations when a sufficient reduction for interactions requires $q > 1$. 
Specifically,  we consider a general case of solving (\ref{LS4}) to approximate the $\bm{X}$-by-$A$ interaction effect term of model (\ref{CSIM.model}). 
Solving the right-hand side of (\ref{LS4}) subject to $\bm{B} \in \Theta_q$ can be viewed as a manifold optimization 
over the space of $p \times q$ matrices subject to the constraint $\bm{B}^\top \bm{B} = \bm{I}_q$, a special case of the Stiefel manifold \citep[see, e.g.,][]{Muirhead}.  
 
 To solve such a constraint optimization problem on a manifold, 
 in this paper, we employ  \texttt{R} \citep{R} package \texttt{ManifoldOptim} \citep{ManifoldOptim} 
 that wraps the \texttt{C++} library \texttt{ROPTLIB} \citep{ROPTLIB}. 
Given a candidate matrix $\bm{B}  = ( \bm{\beta}_1; \ldots;  \bm{\beta}_q)  \in \Theta_q$, 
we can obtain 
an empirical version of the objective function on the right-hand side of (\ref{LS4}), 
analogous to representation 
 (\ref{Q.criterion2}),  
as follows. 
For ease of illustration, 
let us focus on the $q=2$ case. 
For each candidate matrix $\bm{B} \in \Theta_2$, 
at given triplets $(\bm{\beta}_1^\top \bm{x}_i,  \bm{\beta}_2^\top \bm{x}_i, a_i )$ $(i=1,\ldots,n$) 
and two sets of marginal basis 
 $\{ \Psi_r, r=1,\ldots,d \}$ and $\{  \check{\Psi}_s, s= 1,\ldots, \check{d} \}$ 
 associated with $\bm{\beta}_1^\top \bm{X}$ and $\bm{\beta}_2^\top \bm{X}$ respectively, 
the univariate (i.e., $q=1$) basis representation  (\ref{eq.f}) 
 can be extended 
 to a $q(=2)$-dimensional  tensor-product 
 representation: 
 $g_{a_i}(\bm{\beta}_1^\top \bm{x}_i,  \bm{\beta}_2^\top \bm{x}_i)   =\sum_{r=1}^d \sum_{s=1}^{\check{d}} \Psi_r (\bm{\beta}_1^\top \bm{x}_i) \check{\Psi}_s(\bm{\beta}_2^\top \bm{x}_i) \theta_{rs,a_i} 
 = ( \bm{\Psi}_i \otimes \check{\bm{\Psi}}_i ) \bm{\theta}_{a_i}$ 
 $(i=1,\ldots,n)$ 
for some $a$-specific vectors 
 $\bm{\theta}_a \in \mathbb{R}^{d\check{d}}$ $(a \in \mathcal{A})$.  
 If the set $\mathcal{A}$   is a  continuous set (as in Section~\ref{continuous.treatment}),  
   we can allow the coefficients $\bm{\theta}_{a}$   
   to vary smoothly over $a \in \mathcal{A}$, 
   as is assumed in 
  representation  (\ref{tensor.product}).   
  This method of constructing a tensor-product model can be applied to a general
$q > 2$ case. 
Similarly, the associated model penalty matrices 
can be constructed from a set of $q$ roughness penalty matrices of the $q$ individual axes (as in Section~\ref{continuous.treatment}).  
 The linear constraint (\ref{beta.t.condition})  
 (or, that of of type (\ref{lin.constr}), if we work with a continuous set $\mathcal{A}$) can then be absorbed into the tensor product representation of the design and the associated penalty matrices. 
Thus, for each candidate $\bm{B} \in \Theta_2$, 
the penalized least squares criterion of the form  (\ref{Q.criterion2}) (with an appropriate change to the penalty term to penalize over the general $q$ number of  axes) 
can be optimized over $\tilde{\bm{\theta}}$ (with the associated smoothing parameters estimated via,  for example, REML), 
resulting in a profiled objective function over $\bm{B} \in \Theta_2$. 
To optimize over $\bm{B} \in \Theta_q$, 
we can utilize 
a quasi-Newton 
\citep[e.g., BFGS;][]{Fletcher1987} 
method based on numerical approximation to the gradient 
 of the profiled objective function (with respect to $\bm{B}$) via finite differences, as implemented in \texttt{ManifoldOptim} \citep{ManifoldOptim}. 

In practice, the structural dimension $q$ 
of the dimension reduction model (\ref{CSIM.model}) is unknown, and therefore it is viewed as a tuning parameter. 
We next describe how to choose the structural dimension $q$ from data.  
The solution $(g_{01}, \ldots, g_{0L}, \bm{B}_0)$ on the left-hand side of  (\ref{LS4}) is optimal  
 with respect to minimizing the Kullback-Leibler (K-L) divergence between the working 
 model 
$\mathbb{E}[ Y  | \bm{X}, A=a] \approx g_a(\bm{B}^\top \bm{X})$ 
$(g_a \in \mathcal{H}^{(\bm{B})}, \bm{B} \in \Theta_q)$ $(a=1,\ldots,L)$ 
subject to constraint $\mathbb{E}[ g_A(\bm{B}^\top \bm{X}) |\bm{X}]=0$ $(\forall \bm{B} \in \Theta_q)$ 
and the true underlying model (\ref{CSIM.model}) (under Gaussian noise). 
We can utilize an estimate of 
the expected K-L divergence 
of the fitted working model given each value of $q$ based on a cross-validation, 
as a basis of model selection. 
Alternatively, we can utilize the Akaike information criterion \citep[AIC;][]{AIC} based model selection, 
or the network information criterion (NIC) introduced by \cite{NIC},  a generalization of the AIC, in the context of artificial neural networks (ANN).
For NIC, the number of hidden units 
corresponds to 
 $q$ in our case.
 \cite{Davison2003}  provides illustrations of the closeness between AIC and NIC, even when the candidate (i.e., the working) models are incorrectly specified. 
We have  found that, AIC, as a simpler approximation to the (relative) expected K-L divergence  than NIC, 
  behaves closely  to  a cross-validation estimate of the expected K-L divergence,  
and is relatively straightforward to compute. 
In general, AIC is defined to be the negative log likelihood  of the model,    
plus 
two times the (effective) number of parameters used in the model that penalizes  
 the model complexity.
In our setting, 
 the smooths $\{ g_a \}_{a \in \mathcal{A}}$ are represented by  
 a finite dimensional basis $\Psi(\cdot) \in \mathbb{R}^d$ (for $q>1$, we use an appropriate tensor product representation) 
with the  
   associated basis coefficients $\bm{\theta}_a$  penalized for the function smoothness. 
Therefore, to define the AIC penalty term, we  utilize the effective degrees of freedom  \citep{GAM} associated with the basis coefficients  $\bm{\theta}_a$, 
and also account for 
 the smoothing parameter $(\lambda_a)$ estimation uncertainty 
 by the method of \cite{Wood2016}, implemented through  \texttt{R} package \texttt{mgcv} \citep{mgcv}. 
 Let 
$\mbox{AIC}_q^{(g)}$  
denote 
 AIC   associated with the estimated smooths $g_a(\bm{B}^\top \bm{X})$ $(a \in \mathcal{A})$,  for a fixed $\bm{B} \in \Theta_q$. 
Then, we add  to 
$\mbox{AIC}_q^{(g)}$ 
the additional penalty term, $2 q (p-1)$,  associated with the $q (p-1)$  ``free'' parameters of the dimension reduction matrix $\bm{B} \in \Theta_q$, 
to define AIC of the model: 
\begin{equation}  \label{AIC}
\mbox{AIC}_q^{(g)}  + 2 q (p-1), 
\end{equation}
whch can be minimized (over $q$)  
to determine  the structural dimension of model (\ref{CSIM.model}). 

  \section{APPLICATION}  
  
In this section, we apply the concept of sufficient reduction to a dataset from 
a randomized clinical trial for treatment of major depressive depression, 
comparing three (i.e., $L=3$) treatment conditions 
$A=a$ $(a=1,2,3)$:  
$a=1$ corresponds to  placebo; 
$a=2$ corresponds to fluoxetine-varying dose; 
$a=3$ corresponds to  imipramine-varying dose. 
The outcome $Y$ is taken to be 
the improvement in depression symptom severity measured by the Hamilton rating scale for depression (HRSD), 
defined to be HRSD at week 0 - HRSD at week 8,   
and a larger value of $Y$ is desired. 
We consider $p=6$ pretreatment patient characteristics $\bm{X} = (X_1,\ldots, X_6)^\top$: 
baseline symptom severity ($X_1$) ranging from 1 (normal) to 7 (extremely ill); 
age ($X_2$); gender ($X_3$) (0 = female, 1 = male); height ($X_4$); weight ($X_5$); and days of current illness ($X_6$). 
 Each variable is standardized to  zero mean and unit variance. 
 The number of subjects $n=369$.

\begin{figure} 
\begin{center} 
\includegraphics[width=4.2in, height = 1.6in]{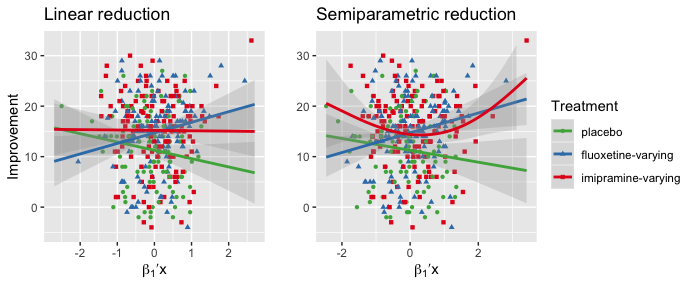} 
\end{center} 
\vspace*{-10pt}
\caption[]{
The scatter plots and the treatment $a$-specific  functions $g_a(\cdot)$ ($a=1,2,3$; i.e., 
placebo, fluoxetine-varying dose and imipramine-varying dose, respectively)  
on the estimated  1-D  reduction  $\bm{\beta}_1^\top \bm{X}$, 
for the linear reduction (left) and the semiparametric reduction (right).    
 } \label{fig3}
\end{figure}

   First, we estimate the  interaction effect part of the (1-D) linear $\bm{X}$-by-$A$ interaction effect model (\ref{eq2.constrained.simml}), 
optimized based on criterion (\ref{least squares2}).  
With $L=3$, there are at most two $(=L-1)$ nonzero eigenvalues associated with the ($6 \times 6$) matrix $\bm{H}$ in (\ref{between.matrix}).
These two eigenvalues are $6.35$ and $0.31$, respectively. 
Compared to the first eigenvalue ($=6.35$, associated with 
  $\bm{\xi}_1$), the second eigenvalue ($=0.31$, associated with  $\bm{\xi}_2$) is relatively negligible. 
This indicates that the 1-D approximation 
    model    (\ref{eq2.constrained.simml}) 
  with $\bm{\beta} = \bm{\xi}_1$  is essentially sufficient for modeling the $\bm{X}$-by-$A$ interaction effects, 
  under the assumption that the linear interaction effect 
  model (\ref{linear.model}) is correctly specified, 
  and therefore, we do not consider a 2-D dimension reduction with the dimension reduction matrix $\bm{B} = ( \bm{\xi}_1; \bm{\xi}_2)$ for this example. 
  The estimated  1-D reduction vector is $\bm{\beta}_1 = (0.24, 0.13,  -0.73,  0.53, -0.08, 0.30)^\top$, and 
the corresponding treatment $a$-specific $(a=1,2,3)$ linear model fits  on the estimated reduction $\bm{\beta}_1^\top \bm{X}$ are illustrated 
in  the 
left panel of Figure~\ref{fig3}.

\begin{figure} 
\begin{center} 
\includegraphics[width=5in, height = 2in]{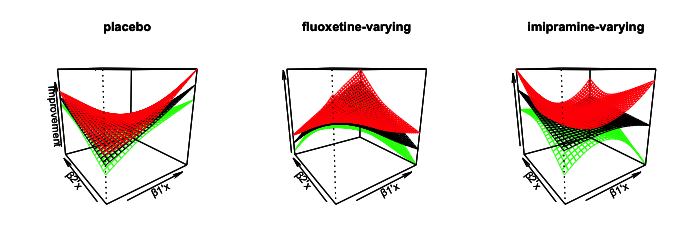} 
\end{center} 
\vspace*{-25pt}
\caption[]{
The treatment $a$-specific  functions $g_a(\cdot)$ ($a=1,2,3$; i.e., 
placebo, fluoxetine-varying dose and imipramine-varying dose, respectively) 
on the estimated  2-dimensional reduction $(\bm{\beta}_1^\top \bm{X}, \bm{\beta}_2^\top \bm{X})$; 
the red and green surfaces are at plus or minus one standard error from the  estimated function (the black surface in the middle) conditional on the estimated  $(\bm{\beta}_1^\top \bm{X}, \bm{\beta}_2^\top \bm{X})$.
 } \label{fig4}
\end{figure}


  
 Second, we estimate the 
  semiparametric dimension reduction model 
   (\ref{CSIM.model}) with $q=1$ (i.e., 1-D reduction), 
  optimized based on criterion (\ref{the.condition2}).  The estimated reduction vector is  
  $\bm{\beta}_1 = (0.42, 0.26,  -0.69,  0.37, -0.14, 0.34)^\top$, and 
  the corresponding treatment $a$-specific ($a=1,2,3)$ curves on the estimated  1-D reduction $\bm{\beta}_1^\top \bm{X}$ 
  are illustrated in the right panel of Figure~\ref{fig3}. 
  The ``imipramine-varying dose'' effect  (i.e., $a=3)$ 
   is clearly better captured by the flexible link function $g_3$ (the red curve)  
as compared to the linear reduction fit illustrated in the middle panel. 
  
  In addition, we estimate the 
  semiparametric dimension reduction model  (\ref{CSIM.model})  
  with $q=2$ (i.e., 2-D reduction),  
  optimized based on criterion (\ref{LS4}). 
 The estimated reduction vectors are  
$\bm{\beta}_1 = (0.56, 0.10,  -0.56,  0.17, 0.34, 0.45)^\top$ 
  and  $\bm{\beta}_2 = (0.29, -0.15,  0.54,  -0.32, -0.23, 0.65)^\top$, and 
  the corresponding treatment $a$-specific ($a=1,2,3)$ surfaces on the estimated 2-D reduction $(\bm{\beta}_1^\top \bm{X}, \bm{\beta}_2^\top \bm{X})$ 
  are illustrated in Figure~\ref{fig4}. 
 
To compare these three estimated dimension reduction models, 
we evaluate AIC (\ref{AIC}). 
The resulting AIC values  are $2529$, $2523$ and $2531$,  
for the 1-D linear,  1-D semiparametric, 
and  2-D semiparametric reduction models, respectively. 
The 1-D semiparametric reduction is favored, with respect to AIC  (\ref{AIC}).  
 Lastly, since this dimension reduction framework  for the $\bm{X}$-by-$A$
interaction effect was 
 motivated from the problem of 
 estimating the optimal individualized treatment rule $\mathcal{D}^{opt}$  in (\ref{dopt}) that maximizes the value $V(\mathcal{D})$, 
we evaluate the performance of  $\hat{\mathcal{D}}^{opt}$,  
with respect to the corresponding value  $V(\hat{\mathcal{D}}^{opt})$, 
where $\hat{\mathcal{D}}^{opt}$  denotes an estimate of $\mathcal{D}^{opt}$ constructed based on each dimension reduction model. 
To estimate the value $V(\hat{\mathcal{D}}^{opt})$, 
we randomly split the dataset (of size $n=369$) at a ratio of $5$ to $1$ into a training set and a testing set (of size $\tilde{n}$), replicated $200$ times, 
each time computing  $\hat{\mathcal{D}}^{opt}$  based on  the training set  and  estimating the  corresponding value $V(\hat{\mathcal{D}}^{opt})$  
by an inverse probability weighted estimator \citep[IPWE;][]{Murphy2005}:  
$\hat{V}(\hat{\mathcal{D}}^{opt}) = \sum_{i=1}^{\tilde{n}} y_{i} 1_{(a_i = \hat{\mathcal{D}}^{opt}(\bm{x}_i))} / \sum_{i=1}^{\tilde{n}}1_{(a_i = \hat{\mathcal{D}}^{opt}( \bm{x}_i))}$  
evaluated based on the testing set.  
The resulting averages (and standard deviations) of 
$\hat{V}(\hat{\mathcal{D}}^{opt})$ 
over the 200 randomly split datasets are 
$14.72 (1.32)$,  $14.77 (1.31)$ and $14.35 (1.37)$, 
 for the 1-D linear,  1-D semiparametric, and 2-D semiparametric reduction models, respectively. 
With respect to  
the value $V(\hat{\mathcal{D}}^{opt})$, 
 the 1-D semiparametric reduction model 
is favored for this dataset. 

\section{DISCUSSION}
Sufficient subspace reductions in regression of $Y$ on $(\bm{X}, A)$ have typically been focused on the main effect of $(\bm{X}, A)$.  In some applications, such as precision medicine, with $A$ representing a treatment variable and $\bm{X}$ representing a set of pretreatment covariates, the primary concern is 
not on the main effect of $\bm{X}$ (which is often considered as a nuisance), but 
on the $\bm{X}$-by-$A$ interactions effect. 
In this paper, we extended the notion of sufficient subspace reduction for the $\bm{X}$ main effects to the $\bm{X}$-by-$A$  interaction effects in regression. 
We  introduced a simple and easy-to-implement optimization framework to estimate a  sufficient subspace for such an interaction effect. 
Linear model-based approaches (e.g., the modified covariate method) and the approach of using a single-index model to estimate the  $\bm{X}$-by-$A$ interaction effects 
are connected 
in  this optimization framework (\ref{LS4}),  
in the context of a randomized clinical trial. This dimension reduction framework does not require to model the $\bm{X}$ main effects when reducing dimension for the $\bm{X}$-by-$A$ interaction effects. Although 
the results in Section~\ref{sec.linear.model} 
rely on the assumption that 
$A$ is distributed independently of $\bm{X}$, 
the general optimization framework   (\ref{LS4}) does not require such an assumption (see, e.g., the implementation of the semiparametric method in Sections~\ref{sec.semipar.model}). 
We also considered an extension of the methods to multiple projections of $\bm{X}$ and to the variable $A$ defined on a continuum.  

One shortcoming of the dimension reduction framework presented in this paper is that the dimension reduction $R(\bm{X})= \bm{B}_0^\top \bm{X}$ for  the $\bm{X}$-by-$A$  interactions
  is defined in terms of all the covariates $\bm{X}$ in the model, i.e., model (\ref{CSIM.model}) forces all the covariates play a role in building an interaction term. 
Also, estimating  (\ref{CSIM.model})   in a high-dimensional $\bm{X}$ space is likely to cause problems of overfitting.
Future work will employ an appropriate  regularization for estimation of a sparse dimension reduction matrix $\bm{B}_0$ 
(subject to  $\bm{B}_0 \in \Theta_q$ for model identifiability), by utilizing, 
 for example, a constrained $L^1$ regularization of \cite{Radchanko}  or 
 a penalized approach of  \cite{Peng.Huang.sim, Wang.Wang.sim}, that can avoid overfitting 
as well as identify important covariates in $\bm{X}$ that 
modify the effects of $A$ on $Y$ as a result of  the $\bm{X}$-by-$A$  interactions. 


\section*{Acknowledgements}

This work was supported by National Institute of Health (NIH) grant 5 R01 MH099003.

\section*{Supporting Information}


\begin{description}
\item[Appendix:] 
The Appendix 
includes the proofs of 
Theorem~\ref{theorem.1}, Corollary~\ref{corollary.1}, Propositions~\ref{proposition.1}, \ref{proposition.2} and \ref{proposition.3}. 

\item[R-packages:] 

For  the case of a 1-D (i.e., $q=1$) reduction, 
the \texttt{R} packages \texttt{simml} \citep[Single-Index Models with Multiple-Links;][]{simml.package} developed for a categorical variable $A$ (described in Section~\ref{sec.semipar.model}) 
and \texttt{simsl} \citep[Single-Index Models with a Surface-Link;][]{simsl.package} developed for a continuous variable $A$ (described in Section~\ref{continuous.treatment}), available on CRAN  \citep{R}, 
provide an implementation of
the proposed dimension reduction method. 

\end{description}

\bibliography{refs}

\begin{thebibliography}{50}
\expandafter\ifx\csname natexlab\endcsname\relax\def\natexlab#1{#1}\fi

\bibitem[Adragni and Cook(2009)]{Adragni2009}
Adragni, K.~P. and Cook, D.~R. (2009).
\newblock Sufficient dimension reduction and prediction in regression.
\newblock \emph{Philosophical Transactions of the Royal Society}
  \textbf{367}:4385--4405.

\bibitem[Adragni \emph{et~al.}(2017)Adragni, Martin, Raim, and
  Huang]{ManifoldOptim}
Adragni, K.~P., Martin, S., Raim, A., and Huang, W. (2017).
\newblock {M}anifold{O}ptim: {A}n {R} interface to the '{ROPTLIB}' library for
  {R}iemannian manifold optimization.
\newblock \emph{R package version 0.1.4} .

\bibitem[Akaike(1974)]{AIC}
Akaike, H. (1974).
\newblock A new look at the statistical model identification.
\newblock \emph{IEEE Transactions on Automatirc Control} \textbf{19}:716--723.

\bibitem[Bura and Cook(2001)]{Bura2001}
Bura, E. and Cook, R.~D. (2001).
\newblock Estimating the structural dimension of regression via parametric
  inverse regression.
\newblock \emph{Journal of Royal Statistical Society: Series B} \textbf{63}.

\bibitem[Cai \emph{et~al.}(2011)Cai, Tian, Wong, and Wei]{Cai}
Cai, T., Tian, L., Wong, P.~H., and Wei, L.~J. (2011).
\newblock Analysis of randomized comparative clinical trial data for
  personalized treatment selections.
\newblock \emph{Biostatistics} \textbf{12}:270--282.

\bibitem[Cook(1998)]{Cook1998a}
Cook, D.~R. (1998).
\newblock \emph{Regression Graphics}.
\newblock Wiley, New York.

\bibitem[Cook and Li(2002)]{Cook2002}
Cook, D.~R. and Li, B. (2002).
\newblock Dimension reduction for conditional mean in regression.
\newblock \emph{The Annals of Statistics} \textbf{30}:455--474.

\bibitem[Cook(1994)]{Cook1994}
Cook, R.~D. (1994).
\newblock On the interpretation of regression plots.
\newblock \emph{Journal of the American Statistical Association}
  \textbf{89}:177--189.

\bibitem[Cook(1996)]{Cook1996}
Cook, R.~D. (1996).
\newblock Graphics for regressions with a binary response.
\newblock \emph{Journal of the American Statistical Association}
  \textbf{91}:983--992.

\bibitem[Cook(2007)]{Cook.dimension.reduction}
Cook, R.~D. (2007).
\newblock Fisher lecture: Dimension reduction in regression.
\newblock \emph{Statistical Science} \textbf{22}:1--26.

\bibitem[Davidson(2003)]{Davison2003}
Davidson, A. (2003).
\newblock \emph{Statistical Models}.
\newblock Cambridge: Cambridge University Press.

\bibitem[de~Boor(2001)]{deBoor}
de~Boor, C. (2001).
\newblock \emph{A Practical Guide to Splines}.
\newblock Springer-Verlag, New York.

\bibitem[Eilers and Marx(1996)]{Eilers1996}
Eilers, P. and Marx, B. (1996).
\newblock Flexible smoothing with {B}-splines and penalties.
\newblock \emph{Statistical Science} \textbf{11}:89--121.

\bibitem[Eilers and Marx(2003)]{PSR.interaction}
Eilers, P. and Marx, B. (2003).
\newblock Multivariate calibration with temperature interaction using
  two-dimensional penalized signal regression.
\newblock \emph{Chemometrics and Intellegence Laboratory Systems}
  \textbf{66}:159--174.

\bibitem[Fisher(1922)]{Fisher1922}
Fisher, R.~A. (1922).
\newblock On the mathematical foundations of theoretical statistics.
\newblock \emph{Philosophical Transactions of the Royal Society}
  \textbf{222}:309--368.

\bibitem[Fletcher(1987)]{Fletcher1987}
Fletcher, R. (1987).
\newblock \emph{Practical Methods of Optimization}.
\newblock Chichester, New York: Wiley.

\bibitem[Hastie and Tibshirani(1999)]{GAM}
Hastie, T. and Tibshirani, R. (1999).
\newblock \emph{Generalized Additive Models}.
\newblock Chapman \& Hall Ltd.

\bibitem[Huang \emph{et~al.}(2016)Huang, Absil, Gallivan, and Hand]{ROPTLIB}
Huang, W., Absil, P., Gallivan, K.~A., and Hand, P. (2016).
\newblock {ROPTLIB}: an object-oriented {C}++ library for optimization on
  {R}iemannian manifolds.
\newblock \emph{Technical Report FSU16-14, Florida State University} .

\bibitem[Jeng \emph{et~al.}(2018)Jeng, Lu, and Peng]{A.learning.Jeng2018}
Jeng, X., Lu, W., and Peng, H. (2018).
\newblock High-dimensional inference for personalized treatment decision.
\newblock \emph{Electronic Journal of Statistics} \textbf{12}:2074--2089.

\bibitem[Li(1991)]{Li1991}
Li, K.-C. (1991).
\newblock Sliced inverse regression for dimension reduction (with discussion).
\newblock \emph{Journal of the American Statistical Association}
  \textbf{86}:316--342.

\bibitem[Li(1992)]{Li1992}
Li, K.-C. (1992).
\newblock On principal {H}essian directions for data visualization and
  dimension reduction: {A}nother application of {S}tein's lemma.
\newblock \emph{Journal of the American Statistical Association}
  \textbf{87}:1025--1039.

\bibitem[Lu \emph{et~al.}(2011)Lu, Zhang, and Zeng]{LU.2011}
Lu, W., Zhang, H., and Zeng, D. (2011).
\newblock Variable selection for optimal treatment decision.
\newblock \emph{Statistical Methods in Medical Research} \textbf{22}:493--504.

\bibitem[Luo \emph{et~al.}(2018)Luo, Wu, and Zhu]{Luo2019}
Luo, W., Wu, W., and Zhu, Y. (2018).
\newblock Learning heterogeneity in causal inference using sufficient dimension
  reduction.
\newblock \emph{Journal of Causal Inference} \textbf{7}.

\bibitem[Luo \emph{et~al.}(2017)Luo, Zhu, and Ghosh]{Luo2017}
Luo, W., Zhu, Y., and Ghosh, D. (2017).
\newblock On estimating regression-based causal effects using sufficient
  dimension reduction.
\newblock \emph{Biometrika} \textbf{104}:51--65.

\bibitem[Marx(2015)]{VCSISR}
Marx, B. (2015).
\newblock Varying-coefficient single-index signal regression.
\newblock \emph{Chemometrics and Intellegence Laboratory Systems}
  \textbf{143}:111--121.

\bibitem[Muirhead(1982)]{Muirhead}
Muirhead, R.~J. (1982).
\newblock \emph{Aspects of Multivariate Statistical Theory}.
\newblock John Wiley \& Sons, Inc., New York.

\bibitem[Murata and Amari(1994)]{NIC}
Murata, N. and Amari, S. (1994).
\newblock {N}etwork {I}nformation {C}riterion- {D}etermining the number of
  hidden units for an artificial neural network model.
\newblock \emph{{IEEE} Transactions on Neural Networks} \textbf{5}:865--872.

\bibitem[Murphy(2003)]{Murphy}
Murphy, S.~A. (2003).
\newblock Optimal dynamic treatment regimes.
\newblock \emph{Journal of the Royal Statistical Society: Series B (Statistical
  Methodology)} \textbf{65}:331--355.

\bibitem[Murphy(2005)]{Murphy2005}
Murphy, S.~A. (2005).
\newblock A generalization error for {Q}-learning.
\newblock \emph{Journal of Machine Learning} \textbf{6}:1073--1097.

\bibitem[Park \emph{et~al.}(2019{\natexlab{a}})Park, Petkova, Tarpey, and
  Ogden]{simml.package}
Park, H., Petkova, E., Tarpey, T., and Ogden, R. (2019{\natexlab{a}}).
\newblock simml: single-index models with multiple-links.
\newblock \emph{R package version 0.1.0} .

\bibitem[Park \emph{et~al.}(2019{\natexlab{b}})Park, Petkova, Tarpey, and
  Ogden]{simsl.package}
Park, H., Petkova, E., Tarpey, T., and Ogden, R. (2019{\natexlab{b}}).
\newblock simsl: single-index models with a surface-link.
\newblock \emph{R package version 0.1.0} .

\bibitem[Park \emph{et~al.}(2020{\natexlab{a}})Park, Petkova, Tarpey, and
  Ogden]{CSIM}
Park, H., Petkova, E., Tarpey, T., and Ogden, R.~T. (2020{\natexlab{a}}).
\newblock A constrained single-index regression for estimating interactions
  between a treatment and covariates.
\newblock \emph{Revision submitted to Biometrics} .

\bibitem[Park \emph{et~al.}(2020{\natexlab{b}})Park, Petkova, Tarpey, and
  Ogden]{SIMML}
Park, H., Petkova, E., Tarpey, T., and Ogden, R.~T. (2020{\natexlab{b}}).
\newblock A single-index model with multiple-links.
\newblock \emph{Journal of Statistical Planning and Inference}
  \textbf{205}:115--128.

\bibitem[Peng and Huang(2011)]{Peng.Huang.sim}
Peng, H. and Huang, T. (2011).
\newblock Penalized least squares for single index models.
\newblock \emph{Journal of Statistical Planning and Inference}
  \textbf{141}:1362--1379.

\bibitem[Petkova \emph{et~al.}(2016)Petkova, Tarpey, Su, and
  Ogden]{GEM.Petkova}
Petkova, E., Tarpey, T., Su, Z., and Ogden, R.~T. (2016).
\newblock Generated effect modifiers in randomized clinical trials.
\newblock \emph{Biostatistics} \textbf{18}:105--118.

\bibitem[Qian and Murphy(2011)]{QianAndMurphy}
Qian, M. and Murphy, S.~A. (2011).
\newblock Performance guarantees for individualized treatment rules.
\newblock \emph{The Annals of Statistics} \textbf{39}:1180--1210.

\bibitem[{R Development Core Team}(2019)]{R}
{R Development Core Team} (2019).
\newblock \emph{R: {A} {L}anguage and {E}nvironment for {S}tatistical
  {C}omputing}.
\newblock {R} {F}oundation for {S}tatistical {C}omputing, Vienna, Austria.

\bibitem[Radchanko(2015)]{Radchanko}
Radchanko, P. (2015).
\newblock High dimensional single index model.
\newblock \emph{Journal of Multivariate Analysis} \textbf{139}:266--282.

\bibitem[Robins(2004)]{Robins}
Robins, J. (2004).
\newblock \emph{Optimal Structural Nested Models for Optimal Sequential
  Decisions}.
\newblock Springer, New York.

\bibitem[Rubin(1974)]{Rubin1974}
Rubin, D. (1974).
\newblock Estimating causal effects of treatments in randomized and
  nonrandomized studies.
\newblock \emph{Journal of Educational Psychology} \textbf{66}:688--701.

\bibitem[Shi \emph{et~al.}(2018)Shi, Fan, Song, and Lu]{A.learning.Shi2018}
Shi, C., Fan, A., Song, R., and Lu, W. (2018).
\newblock High-dimensional {A}-learning for optimal dynamic treatment regimes.
\newblock \emph{The Annals of Statistics} \textbf{46}:925--957.

\bibitem[Shi \emph{et~al.}(2016)Shi, Song, and Lu]{A.learning.Shi2016}
Shi, C., Song, R., and Lu, W. (2016).
\newblock Robust learning for optimal treatment decision with
  np-dimensionality.
\newblock \emph{Electronic Journal of Statistics} \textbf{10}:2894--2921.

\bibitem[Tian \emph{et~al.}(2014)Tian, Alizadeh, Gentles, and Tibshrani]{MCA}
Tian, L., Alizadeh, A., Gentles, A., and Tibshrani, R. (2014).
\newblock A simple method for estimating interactions between a treatment and a
  large number of covariates.
\newblock \emph{Journal of the American Statistical Association}
  \textbf{109}:1517--1532.

\bibitem[Wang and Wang(2015)]{Wang.Wang.sim}
Wang, G. and Wang, L. (2015).
\newblock Spline estimation and variable selection for single-index prediction
  models with diverging number of index parameters.
\newblock \emph{Journal of Statistical Planning and Inference}
  \textbf{162}:1--19.

\bibitem[Wood(2006)]{Wood2006}
Wood, S.~N. (2006).
\newblock Low-rank scale-invariant tensor product smooths for generalized
  additive mixed models.
\newblock \emph{Biometrics} \textbf{62}:1025--1036.

\bibitem[Wood(2017)]{Wood2017}
Wood, S.~N. (2017).
\newblock \emph{Generalized Additive Models: An Introduction with R}.
\newblock Chapman \& Hall/CRC, second edition.

\bibitem[Wood(2019)]{mgcv}
Wood, S.~N. (2019).
\newblock mgcv: Mixed {GAM} computation vehicle with automatic smoothness
  estimation.
\newblock \emph{R package version 1.8.28} .

\bibitem[Wood \emph{et~al.}(2016)Wood, Pya, and Safken]{Wood2016}
Wood, S.~N., Pya, N., and Safken, B. (2016).
\newblock Smoothing parameter and model selection for general smooth models.
\newblock \emph{Journal of the American Statistical Association}
  \textbf{111}:1548--1575.

\bibitem[Yin \emph{et~al.}(2008)Yin, Li, and Cook]{Yin2008}
Yin, X., Li, B., and Cook, D.~R. (2008).
\newblock Successive direction extraction for estimating the central subspace
  in a multiple-index regression.
\newblock \emph{Journal of Multivariate Analysis} \textbf{99}:1733--57.

\bibitem[Zhang \emph{et~al.}(2012)Zhang, Tsiatis, Laber, and
  Davidian]{Zhang.2012}
Zhang, B., Tsiatis, A.~A., Laber, E.~B., and Davidian, M. (2012).
\newblock A robust method for estimating optimal treatment regimes.
\newblock \emph{Biometrics} \textbf{68}:1010--1018.

\end{thebibliography}

\newpage 
\appendix

\section{Appendix}\label{app}

\subsection*{Proof of Theorem~\ref{theorem.1} }

\begin{proof}

 Suppose there is a sufficient  reduction  $R(\bm{X})=\bm{B}_0^\top \bm{X}$ 
  and the associated   unspecified functions 
 $\{g_a\}_{a \in \mathcal{A}}$,  
 i.e., assume  representation (\ref{treatment.contrast.reduction}) (with $\bm{B} = \bm{B}_0$). 
Let $g_{0a}(\bm{B}_0^\top \bm{X}) =  g_a(\bm{B}_0^\top \bm{X})  - \mathbb{E}[g_A(\bm{B}_0^\top \bm{X}) | \bm{X}]$ $(a \in \mathcal{A})$, 
 which, by rearrangement,  gives 
 $g_a(\bm{B}_0^\top \bm{X}) =  \mathbb{E}[g_A(\bm{B}_0^\top \bm{X}) | \bm{X}]  +g_{0a}(\bm{B}_0^\top \bm{X})$ $(a \in \mathcal{A})$, 
where, by definition, the term $\mathbb{E}[g_A(\bm{B}_0^\top \bm{X}) | \bm{X}]$ does not depend on $a$ and the term $g_{0a}(\bm{B}_0^\top \bm{X})$  
$(a \in \mathcal{A})$ 
 satisfies (\ref{the.constraint}). 
Thus, for any contrast vector $\bm{c}$, 
we can rewrite (\ref{treatment.contrast.reduction}) (with $\bm{B} = \bm{B}_0$) as 
  $$
 \sum_{a=1}^L c_a  g_a(\bm{B}_0^\top \bm{X}) =  \sum_{a=1}^L c_a \big\{ \mathbb{E}[g_A(\bm{B}_0^\top \bm{X}) | \bm{X}]  +g_{0a}(\bm{B}_0^\top \bm{X}) \big\} = \sum_{a=1}^L c_a   g_{0a}(\bm{B}_0^\top \bm{X}), 
 $$
 where 
the second equality follows from $\sum_{a=1}^L c_a \mathbb{E}[g_A(\bm{B}_0^\top \bm{X}) | \bm{X}]  =\mathbb{E}[g_A(\bm{B}_0^\top \bm{X}) | \bm{X}]  \sum_{a=1}^L c_a   =  0$. 
Therefore, for representation (\ref{treatment.contrast.reduction}),   
we can always reparametrize the set of functions 
$\{g_a\}_{a \in \mathcal{A}}$ 
by $\{g_{0a}\}_{a \in \mathcal{A}}$ 
 that satisfies (\ref{the.constraint}), implying that we can assume 
 $g_a = g_{0a}$, without loss of generality. 
By  definition (\ref{treatment.contrast}), we can reexpress  (\ref{treatment.contrast.reduction}) (with $\bm{B} = \bm{B}_0$) as 
\begin{equation} \label{thm1.eq1}
\mathcal{C}(\bm{X};  \bm{c}) 
= \sum_{a=1}^L c_a \mathbb{E}\left[ Y  | \bm{X} , A= a \right] 
=  \sum_{a=1}^L c_a   g_{0a}(\bm{B}_0^\top \bm{X}), 
\end{equation}
 for any contrast vector $\bm{c}$. 
 Under the general model (\ref{the.decomposition}), 
 (\ref{thm1.eq1}) 
indicates that the $\bm{X}$-by-$A$ interaction term 
 $g(\bm{X},A=a)$ $(a \in \mathcal{A})$  
  in (\ref{the.decomposition}) corresponds to the term $g_{0a}(\bm{B}_0^\top \bm{X})$ $(a \in \mathcal{A})$ 
   in  (\ref{thm1.eq1}), 
 since (\ref{thm1.eq1}) holds for an arbitrary contrast $\bm{c}= (c_1,\ldots, c_L)$. 
Furthermore, the term 
 $\mu(\bm{X})$ in (\ref{the.decomposition}) corresponds to $\mu(\bm{X})$ of model (\ref{CSIM.model}), 
 since  $\mu(\bm{X})$ of model (\ref{CSIM.model})  represents the unspecified $\bm{X}$ marginal effect. 
Thus, under the general model (\ref{the.decomposition}), 
 (\ref{treatment.contrast.reduction}) implies model  (\ref{CSIM.model}). 

 Conversely, if we assume model (\ref{CSIM.model}), then, 
 by definition (\ref{treatment.contrast}) we have    
 \begin{equation}\label{the.canceling3} 
\mathcal{C}(\bm{X};  \bm{c} )  \ = \ \sum_{a=1}^L c_a   \mathbb{E}\left[ Y | \bm{X}, A=a \right]  =\ 0 \ + \ \sum_{a=1}^L c_a  g_{0a}(\bm{B}_0^\top \bm{X}), 
\end{equation} 
for all contrast vectors $\bm{c}$, where the $\bm{X}$ marginal effect $\mu(\bm{X})$ in (\ref{CSIM.model})  drops out due to 
$\sum_{a=1}^L c_a = 0$. Expression 
(\ref{the.canceling3})  implies that 
 $\bm{B}_0^\top \bm{X}$ is a sufficient reduction for $\mathcal{C}(\bm{X};  \bm{c} )$, 
implying (\ref{treatment.contrast.reduction}) (with $\bm{B} = \bm{B}_0$).  
\end{proof}

\subsection*{Proof of Corollary~\ref{corollary.1}}
\begin{proof}
By Theorem~\ref{theorem.1}, $R(\bm{X}) = \bm{B}_0^\top \bm{X}$ of model   (\ref{CSIM.model}) is a sufficient reduction (\ref{treatment.contrast.reduction}). 
We need to show that $\mbox{span}( \bm{B}_0 )$  is a minimal reduction, and therefore  $\mbox{span}( \bm{B}_0 ) = S_{\mathcal{C}|\bm{X}}$. 
Due to the constraint (\ref{the.constraint}),  $\bm{B}_0$  of model (\ref{CSIM.model}) is not related to the $\bm{X}$ marginal effect, 
therefore there is no ``nuisance'' dimension contained in $\mbox{span}( \bm{B}_0)$. 
Moreover, since $\bm{B}_0 \in \Theta_{q}$, the columns of $\bm{B}_0$ are linearly independent. 
This implies $\bm{B}_0$ is a basis for $S_{\mathcal{C}|\bm{X}}$.

\end{proof}

\subsection*{Proof of Proposition \ref{proposition.1}}
\begin{proof}
Note that $\bm{\eta}_a-\bar{\bm{\eta}} \in \mbox{span}(\bm{\Xi})$ and hence 
$(\bm{\eta}_a-\bar{\bm{\eta}})^\top \bm{X}$ is measurable with respect to $\bm{X}^\top  \bm{\Xi}$.
If model (\ref{linear.model}) holds, then
\begin{eqnarray*}
\mathcal{C}\big(\bm{X}^\top \bm{\Xi}; \bm{c}\big) 
&=&
\sum_{a=1}^L{c_aE[Y \mid \bm{X}^\top \bm{\Xi}, A=a]}\\ 
&=&
\sum_{a=1}^L{c_aE[ E[Y | \bm{X}, A=a]  \mid \bm{X}^\top  \bm{\Xi}, A=a]}\\ 
&=&
\sum_{a=1}^L{c_aE[ \tilde{\mu}(\bm{X}) + \bm{\eta}_a^\top{\bm{X} \mid \bm{X}^\top  \bm{\Xi}, A=a]}}\;\;\mbox{by (\ref{linear.model})}\\ 
&=&
\sum_{a=1}^L{c_aE[ (\bm{\eta}_a-\bar{\bm{\eta}})^\top{\bm{X} \mid \bm{X}^\top  \bm{\Xi}, A=a]}}\;\;\mbox{(by zero-sum constraint on contrast $\bm{c}$)}\\ 
&=&
\sum_{a=1}^L{c_a(\bm{\eta}_a-\bar{\bm{\eta}})^\top \bm{X} }\;\;\mbox{(by the measurability condition)}\\ 
&=&
\sum_{a=1}^L{c_a(\tilde{\mu}(\bm{X}) + \bm{\eta}_a^\top \bm{X} )}\;\;\mbox{(by the zero-sum constraint on contrast $\bm{c}$)}\\ 
&=&
\sum_{a=1}^L{c_a E[Y \mid \bm{X}, A=a]}\;\;\mbox{(by (\ref{linear.model}))}\\ 
&=&
\mathcal{C}\left(\bm{X}; \bm{c}\right). \\
\end{eqnarray*}
That $(\bm{\eta}_1, \ldots, \bm{\eta}_L)$  are distinct and $\pi_a >0$ is sufficient to guarantee that there are $L-1$ nonzero eigenvalues in the matrix $\bm{H}$ in (\ref{between.matrix}). 
Since the ``between'' group dispersion matrix $\bm{H}$ in (\ref{between.matrix}) has $L-1$ nonzero eigenvalues and the rank of $\bm{\Xi}$ is $L-1$, 
it is clear 
$\mbox{span}(\bm{\Xi}) = S_{\mathcal{C}|\bm{X}}$. 

\end{proof}

\subsection*{Proof of  Proposition~\ref{proposition.2}} 

\begin{proof}
Let $Y_a$ denote 
 $Y$ given  $A=a$ $(a=1,\ldots, L)$, 
   i.e., the $a$-specific outcome. 
For a given $\bm{\beta}$, 
consider the expression:
$$\mathbb{E}\left[ (Y  -  \gamma_{A}{\bm{\beta}}^\top  \bm{X})^2  \right] 
=
 \sum_{a=1}^L \pi_a {\mathbb{E} \left[  (Y_a  -  \gamma_{a}{\bm{\beta}}^\top  \bm{X} )^2 1_{(A=a)}  \right] }
 = \sum_{a=1}^L  \pi_a {\mathbb{E} \left[  (Y_a  -  \gamma_{a}{\bm{\beta}}^\top  \bm{X} )^2  \right] }, 
$$
which can be minimized 
by minimizing each of the $L$ terms with respect to
 $\gamma_a$ $(a=1,\dots, L)$ separately. For the uncentered $\tilde{\gamma}_a$, standard least-squares theory gives the solution as 
$$\tilde{\gamma}_a = {\mbox{cov}(\bm{\beta}^\top\bm{X}, Y_a)\over \mbox{var}(\bm{\beta}^\top\bm{X})}=
{\bm{\beta}^\top \mbox{cov}(\bm{X}, Y_a) \over \bm{\beta}^\top \bm{\Sigma}\bm{\beta}} \quad (a=1,\ldots,L).$$
Because $\bm{X}$ is centered and 
$Y_a$ is centered within each treatment $a$, the covariance in the numerator can be written as
$$ \mbox{cov}(\bm{X}, Y_a)=
\mathbb{E} [\bm{X} Y_a] = \mathbb{E} [ \bm{X} \mathbb{E} [Y_a | \bm{X}]]
= \mathbb{E} [ \bm{X} \bm{X}^\top \bm{\eta}_a] =  \mathbb{E} [ \bm{X}\bm{X}^\top] \bm{\eta}_a=  \bm{\Sigma}\bm{\eta}_a,
$$
and hence 
$$\tilde{\gamma}_a = {\bm{\beta}^\top \bm{\Sigma}\bm{\eta}_a \over \bm{\beta}^\top \bm{\Sigma}\bm{\beta}} \quad (a=1,\ldots,L).$$
Centering the $\tilde{\gamma}_a$ finishes the proof.

\end{proof}

\subsection*{Proof of  Proposition~\ref{proposition.3}} 

\begin{proof}
Consider the criterion of (\ref{least squares2}) at the minimum: 
\begin{equation} \label{eq.1a}
\begin{aligned}
 (\ast \ast) &=  \min_{(\gamma_1, \gamma_2, \bm{\beta})} \
  \mathbb{E}[(Y  -  \bm{X}^\top   \bm{\beta} \gamma_A)^2 ]  \\
  &=   \min_{(\gamma_1, \gamma_2, \bm{\beta})} \ 
  \pi_1 \mathbb{E}[ (Y  -  \bm{X}^\top   \bm{\beta} \gamma_1)^2  \mid A=1]  + (1 -\pi_1) \mathbb{E}[(Y  -  \bm{X}^\top   \bm{\beta} \gamma_2)^2  \mid A=2 ] 
\end{aligned}
\end{equation} 
By Theorem~\ref{theorem.2}, 
the minimum $(\ast \ast) $ occurs at $\bm{\beta}=\bm{\xi}_1$ and 
$\gamma_{a} = ( {\bm{\xi}_1}^\top  \bm{\Sigma}  {\bm{\xi}_1})^{-1}  {\bm{\xi}_1}^\top  \bm{\Sigma}   (\bm{\eta}_a - \bar{\bm{\eta}}) 
=  ( {\bm{\xi}_1}^\top  \bm{\Sigma}  \bm{\xi}_1)^{-1}  {\bm{\xi}_1}^\top  \bm{\Sigma}   \left( \bm{\eta}_a - \{ \pi_1 \bm{\eta}_1 +  (1-\pi_1) \bm{\eta}_2 \} \right)$  $(a=1,2)$, that is: 
\begin{equation}   \label{eq.2a}
\begin{aligned}
\gamma_1 &=  ( {\bm{\xi}_1}^\top  \bm{\Sigma}  \bm{\xi}_1)^{-1}  {\bm{\xi}_1}^\top  \bm{\Sigma}   (\bm{\eta}_2 - \bm{\eta}_1 )(\pi_1-1) = \lVert \bm{\eta}_2 - \bm{\eta}_1 \rVert (\pi_1-1) \quad \mbox{and}  \\ 
\gamma_2 &=  ( {\bm{\xi}_1}^\top  \bm{\Sigma}  \bm{\xi}_1)^{-1}  {\bm{\xi}_1}^\top  \bm{\Sigma}  ( \bm{\eta}_2 - \bm{\eta}_1 ) \pi_1   = \lVert \bm{\eta}_2 - \bm{\eta}_1 \rVert  \pi_1, 
\end{aligned}
\end{equation} 
which follows from $\bm{\xi}_1 = (\bm{\eta}_2 - \bm{\eta}_1)/\lVert \bm{\eta}_2 - \bm{\eta}_1 \rVert$.  
Plugging (\ref{eq.2a}) and  $\bm{\beta} (= \bm{\xi}_1) =
 (\bm{\eta}_2 - \bm{\eta}_1)/\lVert \bm{\eta}_2 - \bm{\eta}_1 \rVert$  into the second line of  (\ref{eq.1a}) gives:  
\begin{equation} \label{eq.4a}
\begin{aligned}
(\ast \ast) 
=&\pi_1 \mathbb{E}[ (Y  -  \bm{X}^\top  (\bm{\eta}_2 - \bm{\eta}_1 )(\pi_1-1))^2  \mid A=1]  \ + \ (1 -\pi_1) \mathbb{E}[ (Y  - \bm{X}^\top  ( \bm{\eta}_2 - \bm{\eta}_1) \pi_1)^2  \mid A=2 ]   \\ 
=&\pi_1 \mathbb{E}[ (Y  -  \bm{X}^\top  {\bm{\beta}} (\pi_1-1))^2  \mid A=1]  + (1 -\pi_1) \mathbb{E}[ (Y  -  \bm{X}^\top  {\bm{\beta}} \pi_1)^2  \mid A=2]   \\
=&\pi_1 \mathbb{E}[ (Y  -  \bm{X}^\top {\bm{\beta}} (A+  \pi_1  -2))^2  \mid A=1]  + (1 -\pi_1) \mathbb{E}[ (Y  -  \bm{X}^\top {\bm{\beta}} (A+  \pi_1  -2))^2  \mid A=2 ]  
\\ 
=& \  \mathbb{E}[(Y  -  \bm{X}^\top {\bm{\beta}} (A+  \pi_1  -2))^2 ], 
\end{aligned}
\end{equation} 
in which we set $\bm{\beta} = (\bm{\eta}_2 - \bm{\eta}_1) \in \mathbb{R}^p$. 
The last line of (\ref{eq.4a}) is the least squares criterion  
on the right-hand side of (\ref{the.MC.criterion}) associated with  $ {\bm{\beta}}^\ast$ of model (\ref{the.MC}). 
Since the minimum $(\ast \ast)$  (\ref{eq.1a}) is unique, it follows that 
${\bm{\beta}}^\ast = (\bm{\eta}_2 - \bm{\eta}_1)$, 
which is proportional to $\bm{\xi}_1 =   (\bm{\eta}_2 - \bm{\eta}_1)/ \lVert \bm{\eta}_2 - \bm{\eta}_1 \rVert$. 
\end{proof}

\end{document}